\documentclass[11pt,a4paper]{article}

\usepackage{amsmath,amssymb,tikz,hyperref,empheq}
\usepackage{graphicx}
 \usepackage{verbatim}

\usepackage{tikz}
\usetikzlibrary {decorations.pathmorphing}
\usetikzlibrary {arrows.meta}
\usetikzlibrary{decorations.text}
 \allowdisplaybreaks

\def\be{\begin{equation}}
\def\ee{\end{equation}}
\def\ba#1\ea{\begin{align}#1\end{align}}
\def\bg#1\eg{\begin{gather}#1\end{gather}}
\def\bm#1\em{\begin{multline}#1\end{multline}}
\def\bmd#1\emd{\begin{multlined}#1\end{multlined}}

\def\({\left(}
\def\){\right)}
\def\[{\left[}
\def\]{\right]}

\def \be {\begin{equation}}
\def \ee {\end{equation}}
\def \ba {\begin{array}}
\def \ea {\end{array}}
\def \bea{\begin{eqnarray}}
\def \eea{\end{eqnarray}}

\def\bea{\begin{eqnarray}}
\def\eea{\end{eqnarray}}

\newcommand{\bit}{\begin{itemize}}  \newcommand{\eit}{\end{itemize}}
\newcommand{\ben}{\begin{enumerate}}  \newcommand{\een}{\end{enumerate}}

\long\def\symbolfootnote[#1]#2{\begingroup%
\def\thefootnote{\fnsymbol{footnote}}\footnote[#1]{#2}\endgroup}

\newcommand{\sysu}{{\it School of Physics and Astronomy, Sun Yat-Sen University, 2 Daxue Road, Zhuhai 519082, China}}

\begin{document}
\thispagestyle{empty}
\begin{center}

~\vspace{20pt}

{\Large\bf Massless Entanglement Islands in Cone Holography}

\vspace{25pt}

Dongqi Li, Rong-Xin Miao ${}$\symbolfootnote[1]{Email:~\sf
  miaorx@mail.sysu.edu.cn}

\vspace{10pt}${}$\sysu

\vspace{2cm}

\begin{abstract}
It is controversial whether entanglement islands can exist in massless gravity theories. Recently, it is found that the massless entanglement island appears in wedge holography with DGP gravity on the branes. In this paper, we generalize the discussions to the codim-n holography named cone holography.  For simplicity, we focus on the case with a codim-2 E brane and a codim-1 Q brane. We discuss the effective action, mass spectrum and holographic entanglement entropy for cone holography with DGP terms. We verify that there is massless gravity on the branes, and recover non-trivial entanglement islands and Page curves. Besides, we work out the parameter space which allows entanglement islands and Page curves. 
Compared with wedge holography, there are several new features.  First, one can not add DGP gravity on the codim-2 E brane. That is because the energy density has to be a constant on codim-2 branes for Einstein gravity in bulk. Second, the Hartman-Maldacena surface ends only on the codim-1 Q brane. Third, the Hartman-Maldacena surface can be defined only in a finite time. We notice that this unusual situation also appears in AdS/dCFT and even in AdS/CFT. Fortunately, it does not affect the Page curve since it happens after Page time. Our results provide more support that the entanglement island is consistent with massless gravity theories.
\end{abstract}

\end{center}

\newpage
\setcounter{footnote}{0}
\setcounter{page}{1}

\tableofcontents

\section{Introduction}
Recently, there has been a significant breakthrough in addressing the black hole information paradox \cite{Hawking:1976ra}, where the entanglement islands play a critical role \cite{Penington:2019npb, Almheiri:2019psf, Almheiri:2019hni, Almheiri:2020cfm}. However, it is controversial whether entanglement islands can exist in massless gravity in dimensions higher than two. So far, most discussions of entanglement islands focus on Karch-Randall (KR) braneworld \cite{Karch:2000ct} and AdS/BCFT \cite{Takayanagi:2011zk,Miao:2017gyt,Chu:2017aab,Miao:2018qkc,Chu:2021mvq}, where the gravity on the brane is massive. See \cite{Almheiri:2019psy,Geng:2020qvw,Chen:2020uac,Ling:2020laa} for examples. Besides, \cite{Geng:2020fxl, Geng:2021hlu, Geng:2022fui} find that entanglement islands disappear in a deformed KR braneworld called wedge holography \cite{Akal:2020wfl, Miao:2020oey} with massless gravity on the branes \cite{Hu:2022lxl}. Inspired by the above evidence, \cite{Geng:2021hlu, Geng:2022fui} conjectures that entanglement islands can exist only in massive gravity theories. They argue that the entanglement island is inconsistent with long-range gravity obeying gravitational Gauss's law. However, there are controversies on this conjecture \cite{Krishnan:2020fer, Ghosh:2021axl, Yadav:2022mnv}. Naturally, the general arguments of the island mechanism apply to massless gravity \cite{Almheiri:2020cfm}. Recently, \cite{Miao:2022mdx, Miao:2023unv} recovers massless entanglement islands in wedge holography with Dvali-Gabadadze-Porrati (DGP) gravity \cite{Dvali:2000hr} on the branes. In particular, \cite{Miao:2023unv} discusses an inspiring analog of the island puzzle in AdS/CFT and argues that the island puzzle in wedge holography can be resolved similarly as in AdS/CFT. The results of \cite{Miao:2022mdx, Miao:2023unv} strongly support that entanglement islands are consistent with massless gravity theories. See also \cite{Emparan:2023dxm, Bahiru:2023zlc} for some related works. Interestingly, \cite{Emparan:2023dxm} observes that the absence-of-island issue can be alleviated in the large $D$ limit. Remarkably, \cite{Bahiru:2023zlc} finds that the massless island puzzle can be resolved, provided that the bulk state breaks all asymptotic symmetries. See also
\cite{Rozali:2019day,Chen:2019uhq,Almheiri:2019yqk,Kusuki:2019hcg,
   Balasubramanian:2020hfs,  Kawabata:2021hac,Bhattacharya:2021jrn,Kawabata:2021vyo,Chen:2020hmv,Bhattacharya:2021nqj,Geng:2021mic,Chou:2021boq,Ahn:2021chg,Alishahiha:2020qza,Gan:2022jay,Omidi:2021opl,Hu:2022ymx,Azarnia:2021uch,Anous:2022wqh,Saha:2021ohr,Geng:2022slq,Geng:2022tfc,Yu:2022xlh,Chu:2022ieq,Hu:2022zgy,Yadav:2023qfg,Piao:2023vgm,RoyChowdhury:2022awr,Choudhury:2020hil,Hung:2023mbw,Afrasiar:2023jrj,Perez-Pardavila:2023rdz,Basu:2022crn,Kanda:2023zse} for some recent works on entanglement islands, Page curve and AdS/BCFT.

In this paper, we generalize the discussions of \cite{Miao:2022mdx, Miao:2023unv} to cone holography \cite{Miao:2021ual}.   For simplicity, we focus on the case with a codim-2 $E$ brane and a codim-1 $Q$ brane. Cone holography can be regarded as holographic dual of the edge modes on the codim-n defect, which is a generalization of wedge holography. Remarkably, there is also massless gravity on the branes of cone holography \cite{Miao:2021ual}.   We investigate the effective action, mass spectrum, holographic entanglement entropy and recover entanglement islands and Page curves in cone holography with DGP terms. Compared with wedge holography, there are several new features.  First, one can not add DGP gravity on the codim-2 $E$ brane, since the energy density has to be a constant on codim-2 branes for Einstein gravity in bulk \cite{Bostock:2003cv}. To allow DGP gravity on the codim-2 brane, we can consider Gauss-Bonnet gravity in bulk \cite{Bostock:2003cv}. Second, the Hartman-Maldacena surface ends only on the codim-1 $Q$ brane. Third, the Hartman-Maldacena surface can be defined only in a finite time. Note that this unusual situation also appears in AdS/dCFT  \cite{Hu:2022zgy} and even in AdS/CFT. Fortunately, it does not affect the Page curve since it happens after Page time. Our results provide more support that the entanglement island is consistent with massless gravity theories.

The paper is organized as follows. 
In section 2, we formulate cone holography with DGP gravity on the brane. Then, we find massless gravity on the branes and get a lower bound of the DGP parameter from the holographic entanglement entropy. Section 3 discusses the entanglement island and the Page curve on tensionless codim-2 branes. Section 4 generalizes the discussions to tensive codim-2 branes. Finally, we conclude with some open problems in section 5.

\section{Cone holography with DGP terms}

This section investigates the cone holography with DGP gravity on the brane. First, we work out the effective action for one class of solutions and obtain a lower bound of the DGP parameter to have a positive effective Newton's constant. Second, we analyze the mass spectrum and verify that it includes a massless mode. Third, we calculate the holographic entanglement entropy for a disk and get another lower bound of the DGP parameter.

 \begin{figure}[t]
\centering
\includegraphics[width=12cm]{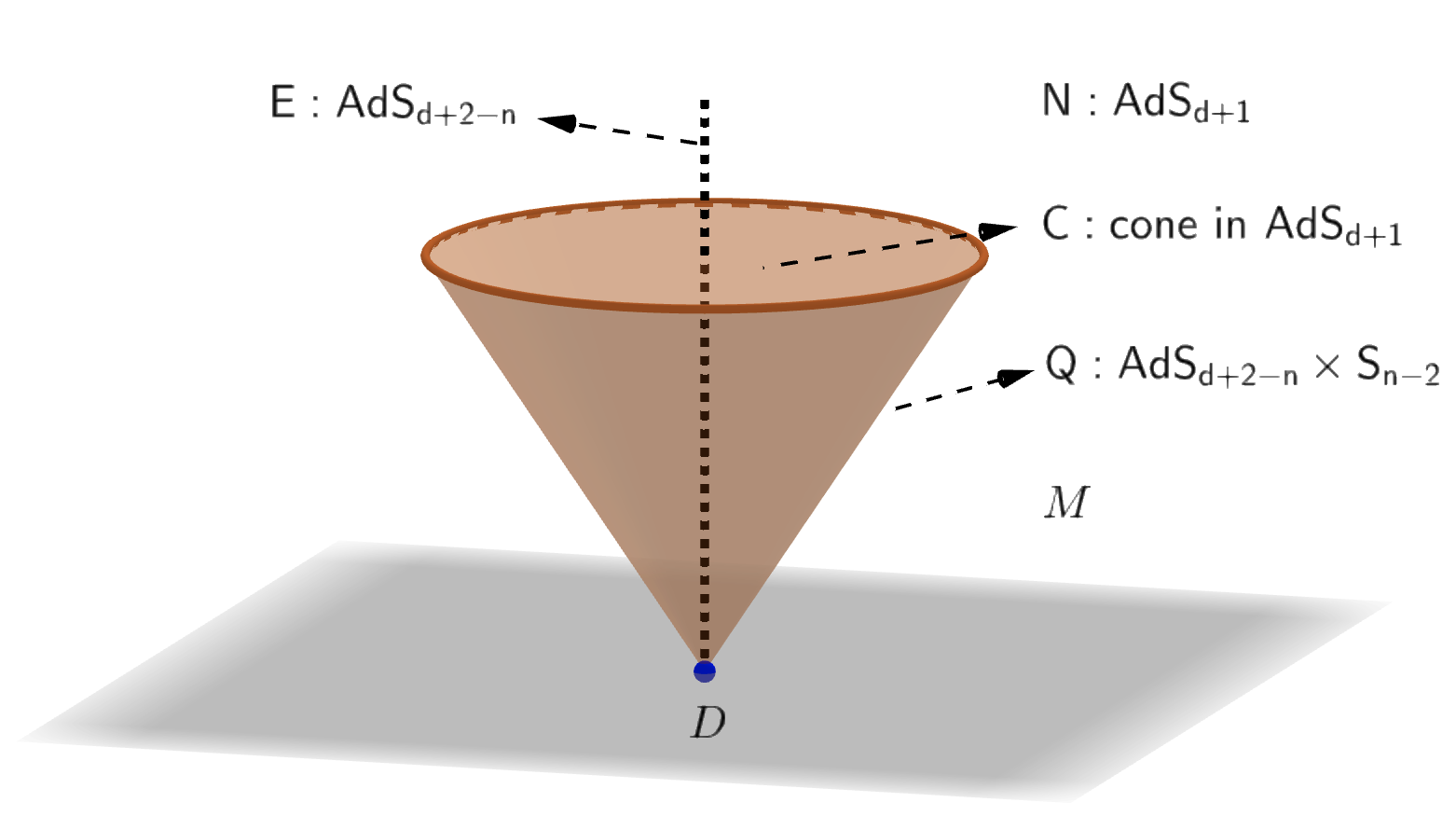}
\caption{ Geometry of cone holography:  $Q$ is a codim-1 brane, $C$ is the cone bounded by $Q$, i.e., $\partial C=Q$, and  $E$ (black dotted line) is a codim-$m$ brane in bulk, where $m=n-1$. The geometries of $Q$ and $E$ are set to be $\text{AdS}_{d+2-n}\times \text{S}_{n-2}$ and $\text{AdS}_{d+2-n}$ so that they shrink to the same defect $D=\partial Q=\partial E$ on the AdS boundary $M$.}
\label{coneholography}
\end{figure}

\begin{figure}[t]
\centering
\includegraphics[width=12cm]{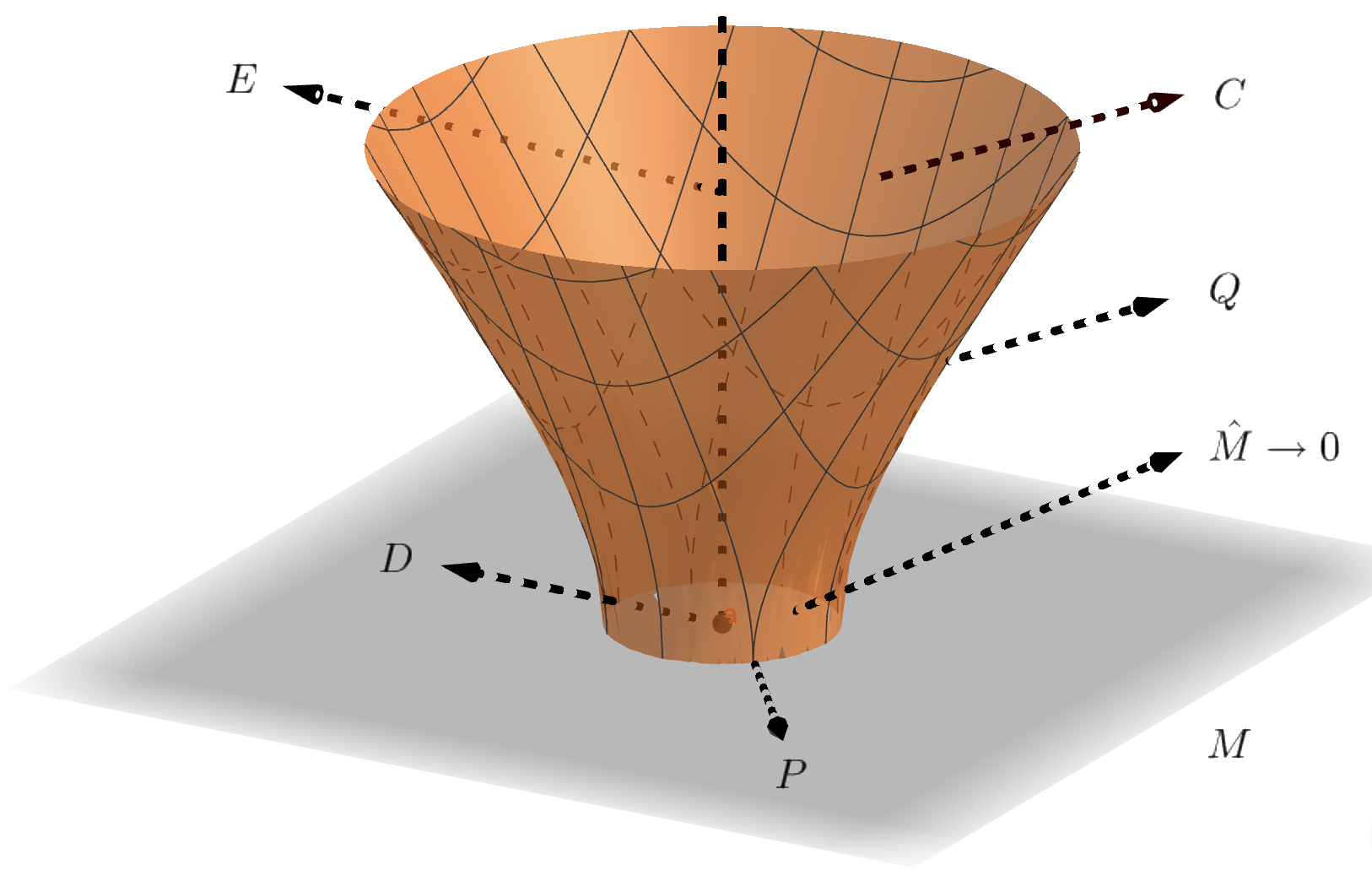}
\caption{Cone holography from AdS/dCFT.  dCFT lives in the manifold $\hat{M}$ with a boundary $P$ and a codim-$m$ defect $D$ at the center. The boundary $P$ and codim-$m$ defect $D$ are extended to an end-of-world brane $Q$ and a codim-$m$  brane $E$ in the bulk, respectively. $C$ (orange) is the bulk spacetime bounded by $Q$ and $\hat{M}$, $M$ (gray) is the AdS boundary. In the limit $\hat{M}\to 0$, the bulk spacetime $C$ becomes a cone and we obtain the cone holography from AdS/dCFT.}
\label{coneholographyfromAdSdCFT}
\end{figure}

Let us illustrate the geometry of cone holography. See Fig.\ref{coneholography}, where $E$ denotes the codim-m brane, $Q$ indicates the codim-1 brane, $C$ is the bulk cone bounded by $Q$, and $D=\partial E=\partial Q$ is the codim-m defect on the AdS boundary $M$. Cone holography proposes that the classical gravity in the bulk cone $C$ is dual to ``quantum gravity" on the branes $E$ and $Q$ and is dual to the CFTs on the defect $D$. Cone holography can be derived from AdS/dCFT by taking the zero volume limit $\hat{M}\to 0$. See Fig. \ref{coneholographyfromAdSdCFT}. In the zero volume limit, the bulk modes disappear, and only the edge modes on the defect survive. Thus cone holography can be regarded as a holographic dual of the edge modes on the defect. For simplicity, we focus on codim-2 brane $E$ in this paper. 

Let us take a typical metric to explain the geometry,
\begin{eqnarray}\label{metric}
ds^2=dr^2+\sinh^2(r)d\theta^2+\cosh^2(r)\frac{dz^2-dt^2+\sum_{\hat{i}=1}^{d-3} dy^2_{\hat{i}}}{z^2}, \quad 0\le r\le \rho,
\end{eqnarray}
where codim-2 brane $E$, codim-1 brane $Q$, and the defect $D$ locate at $r=0$, $r=\rho$ and $z=0$, respectively. 

The action of cone holography with DGP gravity on the brane is given by
\begin{eqnarray}\label{action}
I=\int_C d^{d+1}x\sqrt{-g}\Big(R_C+d(d-1)\Big)-T_E \int_E d^{d-1}x\sqrt{-h_E}+2\int_{Q} d^dx\sqrt{-h_Q}(K-T+\lambda R_{Q}),
\end{eqnarray}
where we have set Newton's constant $16\pi G_N=1$ together with the AdS radius $L=1$, $R_C$ is the Ricci scalar in bulk, $T_E$, $T$ and $\lambda$ are free parameters, $K$, $h_{Q\ ij}$ and $R_{Q}$ are the extrinsic curvature, induced metric, and the intrinsic Ricci scalar (DGP gravity) on the codim-1 brane $Q$, respectively. Note that one cannot add DGP gravity on the codim-2 brane $E$. That is because the energy density has to be a constant on codim-2 branes for Einstein gravity in bulk \cite{Bostock:2003cv}. To allow DGP gravity on codim-2 branes, one can consider higher derivative gravity such as Gauss-Bonnet gravity in bulk \cite{Bostock:2003cv}. 

Recall that the geometry of $Q$ is $\text{AdS}_{d-1}\times \text{S}_{1}$. 
Following \cite{Miao:2021ual}, we choose Dirichlet boundary condition (DBC) on $\text{S}_{1}$ and Neumann boundary condition (NBC) on $\text{AdS}_{d-1}$
\begin{eqnarray}\label{MBC1}
&&\text{DBC}:\ \delta g_{\theta \theta}=0, \\
&&\text{NBC}: K^{ij}-(K-T+\lambda R_{Q}) h_Q^{ij}+2 \lambda R^{ij}_Q=0. \label{MBC2}
\end{eqnarray}
The above boundary condition has the advantage that it is much easier to be solved \cite{Miao:2021ual}. For simplicity, we focus on mixed boundary conditions in this paper. See \cite{Miao:2021ual} for some discussions on the Neumann boundary condition.

\subsection{Effective action}

Now let us discuss the effective action on the branes.  To warm up, we first study the case with tensionless brane $E$, i.e., $T_E=0$.  For simplicity, we focus on the following metric
\begin{eqnarray}\label{metricTE0}
ds^2=dr^2+\sinh^2(r)d\theta^2+\cosh^2(r) h_{ij}(y) dy^i dy^j,\qquad 0\le r\le \rho,
\end{eqnarray}
where $E$ is at $r=0$, $Q$ is at $r=\rho$, $h_{ij}=h_{Q ij}/\cosh^2(\rho)$ obey Einstein equation on the brane $E$
\begin{eqnarray}\label{EOMh}
R_{h\ ij}-\frac{R_{h}+(d-2)(d-3)}{2}h_{ij}=0. 
\end{eqnarray}
The solution (\ref{metricTE0}) obeys the mixed BC (\ref{MBC1},\ref{MBC2}) provided that the parameters are related by
\begin{eqnarray}\label{Tlambda}
T=\coth (\rho )+(d-2) \tanh(\rho)-\lambda \ \text{sech}^2(\rho ) (d-2)(d-3).
\end{eqnarray}
Substituting (\ref{metricTE0}) into the action (\ref{action}) and integrating along $r$ and $\theta$, we obtain the effective action
\begin{eqnarray}\label{AdSCaction1}
I_{\text{eff}}
= \frac{1}{16\pi G^{(d-1)}_{\text{eff}}}\int d^{d-1}y \sqrt{-h} \Big{(} R_{h} +(d-2)(d-3), \Big{)}\end{eqnarray}
with effective Newton's constant
\begin{eqnarray}\label{Newton's constant1}
\frac{1}{16\pi G^{(d-1)}_{\text{eff}}}=2\pi\left( \int_0^{\rho} \sinh(r)\cosh^{d-3}(r) dr+2\lambda \sinh(\rho) \cosh^{d-3}(\rho)\right).
\end{eqnarray}

Let us go on to study the tensive case, i.e., $T_E>0$.  The typical metric is given by  \cite{Miao:2021ual}
\begin{eqnarray}\label{rbarmetric}
ds^2=\frac{d\bar{r}^2}{F(\bar{r})}+ F(\bar{r}) d\theta^2+\bar{r}^2 h_{ij}(y) dy^i dy^j,\qquad \bar{r}_h\le \bar{r}\le \bar{r}_0,
\end{eqnarray}
where $dr=\frac{d\bar{r}}{\sqrt{F(\bar{r})}}$, $F(\bar{r})=\bar{r}^2-1-\frac{\bar{r}_h^{d-2}(\bar{r}_h^{2}-1)}{\bar{r}^{d-2}}$ and $\bar{r}_h=\frac{1+\sqrt{d^2 q^2-2 d q^2+1}}{d q}.$ Note that the codim-2 brane $E$ locates at $\bar{r}=\bar{r}_h$ and the codim-1 brane $Q$ is at $\bar{r}=\bar{r}_0> \bar{r}_h$. The codim-2 brane tension is related to the conical defect
\begin{eqnarray}\label{tensionE}
8\pi G_N T_E= 2\pi \left(1-\frac{1}{q}\right),
\end{eqnarray}
where $2\pi q$ denotes the period of angle $\theta$.  The metric obeys the mixed BC(\ref{MBC1},\ref{MBC2}) provided that we choose the parameters
\begin{eqnarray}\label{Tlambdatensive}
T=\frac{F'(\bar{r}_0)}{2\sqrt{F(\bar{r}_0)}}+(d-2) \frac{\sqrt{F(\bar{r}_0)}}{\bar{r}_0}-\lambda \  \frac{(d-2)(d-3)}{\bar{r}_0^2}.
\end{eqnarray}
One can check that (\ref{Tlambdatensive}) agrees with the tensionless case (\ref{Tlambda}) with $q=1, F(r)=r^2-1, \bar{r}_0=\cosh(\rho)$.  Following the approach of \cite{Miao:2021ual}, we obtain the effective action (\ref{AdSCaction1}) with the effective 
Newton's constant
\begin{eqnarray}\label{Newton's constant2}
\frac{1}{16\pi G^{(d-1)}_{\text{eff}}}=2\pi q\left(\frac{\bar{r}_0^{d-2}-\bar{r}_h^{d-2}}{d-2} +2\lambda \sqrt{F(\bar{r}_0)} \bar{r}_0^{d-3} \right).
\end{eqnarray}

\begin{figure}[t]
\centering
\includegraphics[width=12cm]{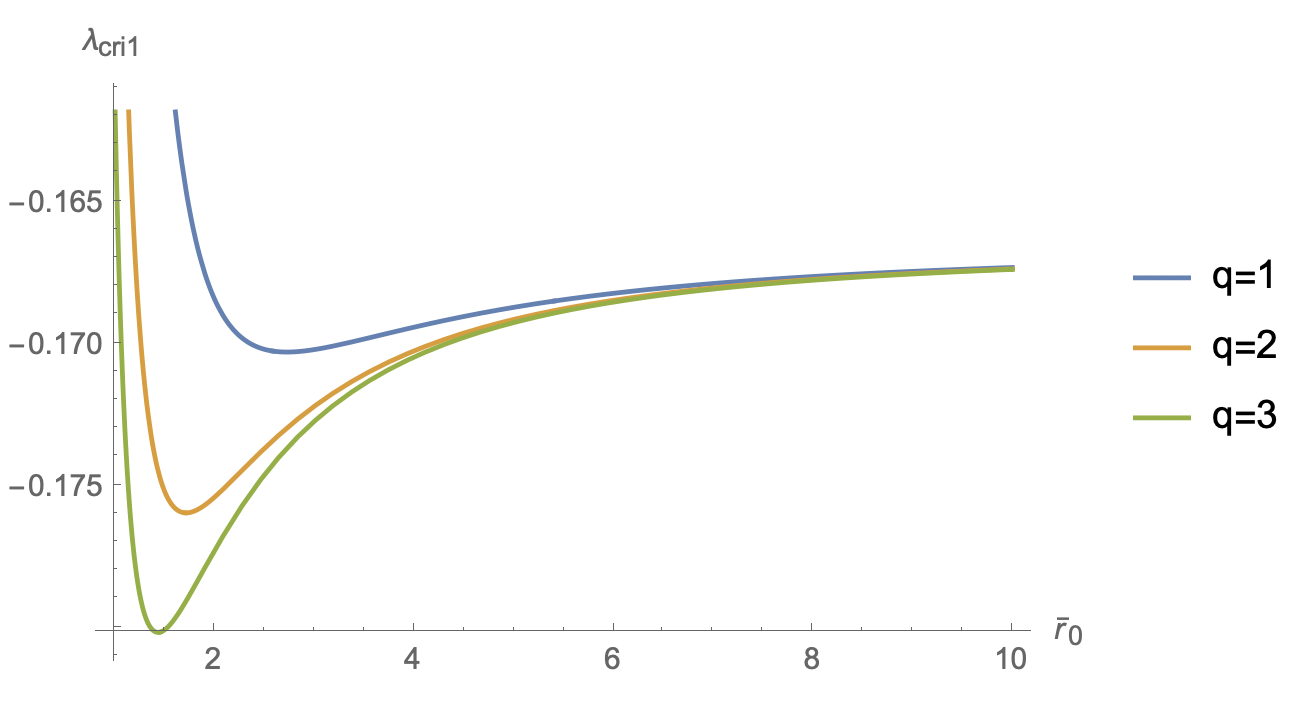}
\caption{The lower bound of DGP parameter for $d=5$. The larger the tension $q$ is, the smaller the lower bound $ \lambda_{\text{cri1}}$ is. In the large $\bar{r}_0$ limit, we have $ \lambda_{\text{cri1}} \to -1/(2(d-2))$. }
\label{lowerbound1}
\end{figure}

Let us make some comments. First, from the effective action (\ref{AdSCaction1}) and EOM (\ref{EOMh}), it is clear that there is massless gravity on the branes. Second, we require that the effective Newton's constant (\ref{Newton's constant2}) is positive, which yields a lower bound on the DGP parameter
\begin{eqnarray}\label{boundDGP1}
\lambda \ge \lambda_{\text{cri1}}=\frac{\bar{r}_0^{3-d} \bar{r}_h^{d-2}-\bar{r}_0}{2 (d-2) \sqrt{F\left(\bar{r}_0\right)}}.
\end{eqnarray}
In the large $\bar{r}_0$ limit, we have $ \lambda_{\text{cri1}} \to -1/(2(d-2))$.
 See Fig.\ref{lowerbound1} for the $\bar{r}_0$ dependence of $ \lambda_{\text{cri1}}$ for $d=5$ and $q=1,2,3$, where $q$ labels the tension (\ref{tensionE}). It shows that the larger the tension $q$ is, the smaller $ \lambda_{\text{cri1}}$ is. 

\subsection{Mass spectrum}

In this subsection, we study the mass spectrum of gravitons for cone holography with DGP gravity on the brane. We find the mass spectrum includes a massless mode, which agrees with the results of the last subsection.  The smaller the DGP parameter is, the larger the mass gap is, the well Einstein gravity behaves as an effective theory at low energy scale.

We first discuss the tensionless case, i.e., $T_E=0$.  
We take the following ansatz of the perturbation metric 
\begin{eqnarray}\label{perturbationmetricTE0}
ds^2=dr^2+\sinh^2(r) d\theta^2+\cosh^2 (r) \left( h^{(0)}_{ij}(y) + H(r) h^{(1)}_{ij}(y)  \right)dy^i dy^j,
\end{eqnarray}
where $h^{(0)}_{ij}(y)$ is the AdS metric with a unit radius and $h^{(1)}_{ij}(y)$ denotes the perturbation.  Note that the above ansatz automatically obeys DBC (\ref{MBC1}) on the $S_1$ sector of the codim-1 brane $Q$.  We impose the transverse traceless gauge 
 \begin{eqnarray}\label{hij1gauge}
D^i h^{(1)}_{ij}=0,\ \ \  h^{(0)ij}h^{(1)}_{ij}=0,
\end{eqnarray}
where $D_i$ is the covariant derivative defined by $h^{(0)}_{ij}$. 
Substituting (\ref{perturbationmetricTE0}) together with (\ref{hij1gauge})  into Einstein equations and separating variables, we obtain
 \begin{eqnarray}\label{EOMMBCmassivehij}
&& \left(D_i D^i+2-m^2\right) h^{(1)}_{ij}(y)=0,\\
&& \sinh (2 r) H''(r)+(d \cosh (2 r)-d+2)H'(r) +2 m^2  \tanh (r) H(r)=0, \label{EOMMBCmassiveH}
\end{eqnarray}
where $m$ labels the mass of gravitons.  Solving (\ref{EOMMBCmassiveH}), we obtain \cite{Miao:2021ual}
 \begin{eqnarray}\label{EOMMBCmassiveHsolution}
H(r)=c_1 \, _2F_1\left(a_1,a_2;1;\tanh ^2(r)\right)+c_2 G_{2,2}^{2,0}\left(\tanh ^2(r)|
\begin{array}{c}
 a_1+\frac{d}{2}, a_2+\frac{d}{2} \\
 0,0 \\
\end{array}
\right),
\end{eqnarray}
where $_2F_1$ is the hypergeometric function, $G_{2,2}^{2,0}$ is the Meijer G function, $c_1$ and $c_2$ are integral constants and $a_i$ are given by
 \begin{eqnarray}\label{aibia1}
&&a_1=\frac{1}{4} \left(2-d-\sqrt{(d-2)^2+4 m^2}\right),\\ \label{aibia2}
&&a_2=\frac{1}{4} \left(2-d+\sqrt{(d-2)^2+4m^2}\right). 
\end{eqnarray}

\begin{figure}[t]
\centering
\includegraphics[width=12cm]{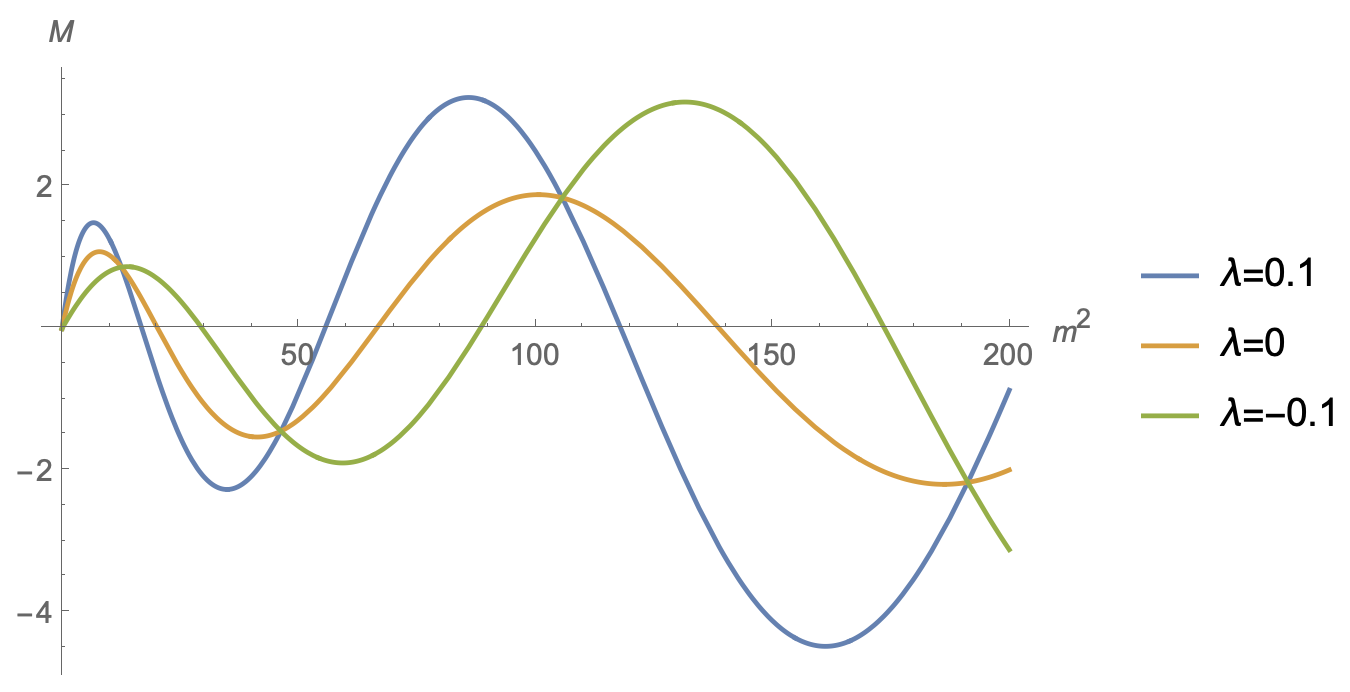}
\caption{ The mass spectrum $M$ (\ref{spectrumTE0}) for $T_E=0, \rho=1$ and $d=5$, where the intersections of the curves and $m^2-$axis denote the allowed mass. The blue, orange and green curves correspond to the DGP parameters $\lambda=0.1, 0, -0.1$, respectively. It includes a massless mode, and the smaller the parameter DGP $\lambda$ is, the larger the mass and mass gap are. }
\label{mass0}
\end{figure}

We choose the natural boundary condition on the codim-2 brane $E$ 
\begin{eqnarray}\label{naturalBCHI}
H(0) \ \text{is finite},
\end{eqnarray}
which yields $c_2=0$ due to the fact $G_{2,2}^{2,0}\left(\tanh ^2(r)|
\begin{array}{c}
 a_1+\frac{d}{2}, a_2+\frac{d}{2} \\
 0,0 \\
\end{array}
\right) \sim \ln r$ for $r\sim 0$.  We impose NBC (\ref{MBC2}) on the $AdS_{d-1}$ sector of the codim-1 brane $Q$
 \begin{eqnarray}\label{MBCH}
\cosh ^2\left(\rho \right) H'\left(\rho \right)-2 \lambda  m^2 H\left(\rho \right)=0,
\end{eqnarray}
where we have used EOM (\ref{EOMMBCmassivehij}) to simplify the above equation.  Substituting the solution (\ref{EOMMBCmassiveHsolution}) with $c_2=0$ into the boundary condition (\ref{MBCH}), we obtain a constraint for the mass spectrum
 \begin{eqnarray}\label{spectrumTE0}
M=\frac{m^2 }{2}\Big(4 \lambda  \, _2F_1\left(a_1,a_2;1;\tanh ^2(\rho )\right)+\tanh (\rho ) \, _2F_1\left(a_1+1,a_2+1;2;\tanh ^2(\rho )\right)\Big)=0
\end{eqnarray}
with $a_1, a_2$ given by (\ref{aibia1},\ref{aibia2}). The mass spectrum (\ref{spectrumTE0}) includes a massless mode $m^2=0$, which agrees with the results of the last subsection. There is an easier way to see this. Clearly, $H(r)=1$ and $m^2=0$ are solutions to EOM (\ref{EOMMBCmassiveH}) and BC (\ref{MBCH}). Furthermore, this massless mode is normalizable
 \begin{eqnarray}\label{sect2:normalizable}
\int_{0}^{\rho} dr \sinh(r)\cosh^{d-3}(r) H(r)^2\ \text{is finite}. 
\end{eqnarray}
Thus, there is indeed a physical massless gravity on the codim-2 brane $E$ in cone holography with DGP gravity. On the other hand, the massless mode is non-normalizable due to the infinite volume in the usual AdS/dCFT \cite{Hu:2022zgy}
 \begin{eqnarray}\label{sect2:non-normalizable}
\int_{0}^{\infty} dr \sinh(r)\cosh^{d-3}(r) H(r)^2 \to \infty. 
\end{eqnarray}
Let us draw the mass spectrum $M(m^2)$ in Fig. {\ref{mass0}}, which shows that there is a massless mode and the smaller the parameter 
DGP $\lambda$ is, the larger the mass and mass gap are. 

Let us go on to discuss the spectrum for tensive codim-2 branes, i.e., $T_E>0$.  We choose the following metric ansatz 
\begin{eqnarray}\label{rbarmetric2}
ds^2=\frac{d\bar{r}^2}{F(\bar{r})}+ F(\bar{r}) d\theta^2+\bar{r}^2  \left( h^{(0)}_{ij}(y) + H(\bar{r}) h^{(1)}_{ij}(y)  \right)dy^i dy^j,\qquad \bar{r}_h\le \bar{r}\le \bar{r}_0,
\end{eqnarray}
where $F(\bar{r})=\bar{r}^2-1-\frac{\bar{r}_h^{d-2}(\bar{r}_h^{2}-1)}{\bar{r}^{d-2}}$ and $\bar{r}_h=\frac{1+\sqrt{d^2 q^2-2 d q^2+1}}{d q}.$ Following the above approaches, we derive the EOM
\begin{eqnarray}\label{EOMHrbar}
H''\left(\bar{r}\right)+\left(\frac{d-1}{\bar{r}}+\frac{F'\left(\bar{r}\right)}{F\left(\bar{r}\right)}\right)H'\left(\bar{r}\right) +\frac{m^2 }{\bar{r}^2 F\left(\bar{r}\right)}H\left(\bar{r}\right)=0,
\end{eqnarray}
and BCs for $H(\bar{r})$
\begin{eqnarray}\label{BCTE1}
&&H(\bar{r}_h) \ \text{is finite}, \\
&&2 \lambda  m^2 H\left(\bar{r}_0\right)-\bar{r}_0^2 \sqrt{F\left(\bar{r}_0\right)} H'\left(\bar{r}_0\right)=0. \label{BCTE2}
\end{eqnarray}
Following the shooting method of \cite{Hu:2022zgy}, we can calculate the mass spectrum numerically. Without loss of generality, we take $d=5, \bar{r}_0=10, \lambda=0.1,0,-0.1$ as examples. We list the mass spectrum for $q=1, 5$ in Table. \ref{table1spectrum} and Table. \ref{table2spectrum}, respectively. Here $q$ labels the tension $T_E$ (\ref{tensionE}), and $q=1$ corresponds to the tensionless case $T_E=0$. Table. \ref{table1spectrum} and Table. \ref{table2spectrum} shows that there is a massless mode, and the mass decreases with the ``tension" $q$ and the DGP parameter $\lambda$.

\begin{table}[ht]
\caption{Mass spectrum for $d=5$ and $q=1$}
\begin{center}
    \begin{tabular}{| c | c | c | c |  c | c | c | c| c| c|c| }
    \hline
     & $1$ & $2$ & 3  & 4& 5 \\ \hline
  $m^2$ for $\lambda=0.1$   & 0 & 10.032 & 28.204 & 54.673& 89.595\\ \hline
  $m^2$ for $\lambda=0$   & 0 & 10.050 & 28.316 & 55.016& 90.353 \\ \hline
  $m^2$ for $\lambda=-0.1$  & 0 & 10.119 & 28.714 & 56.160& 92.718\\ \hline
    \end{tabular}
\end{center}
\label{table1spectrum}
\end{table}

\begin{table}[ht]
\caption{Mass spectrum for $d=5$ and $q=5$}
\begin{center}
    \begin{tabular}{| c | c | c | c |  c | c | c | c| c| c|c| }
    \hline
     & $1$ & $2$ & 3  & 4& 5 \\ \hline
  $m^2$ for $\lambda=0.1$   & 0 & 3.636 & 10.174 & 19.719& 32.251 \\ \hline
  $m^2$ for $\lambda=0$   & 0 & 3.637 & 10.184 & 19.754& 32.334 \\ \hline
  $m^2$ for $\lambda=-0.1$  & 0 & 3.644 & 10.225 & 19.880& 32.623  \\ \hline
    \end{tabular}
\end{center}
\label{table2spectrum}
\end{table}

\subsection{Holographic entanglement entropy}

In this subsection, we study the holographic entanglement entropy (HEE) \cite{Ryu:2006bv} in cone holography with DGP gravity. We discuss HEE for the whole space and a disk subspace on the defect and  obtain another lower bound of the DGP parameter in order to have non-negative HEE.  From the action (\ref{action}), we read off HEE
\begin{eqnarray}\label{HEE}
S_{\text{HEE}}={\text{min}} \left\{ \text{ext} \Big( 4\pi \int_{\Gamma} d^{d-1}x \sqrt{\gamma}+8\pi \int_{\partial \Gamma} d^{d-2}x \sqrt{\sigma} \lambda \Big)\right\},
\end{eqnarray}
where $\Gamma$ denote the RT surface, $\partial \Gamma=\Gamma\cap Q$ is the intersection of the RT surface and the codim-1 brane, $\gamma$ and $\sigma$ represent the induced metric on $\Gamma$ and $\partial \Gamma$ respectively.  For simplicity, we focus on an AdS space in bulk, which means the CFT on the defect is in vacuum. 

Let us comment on how to derive HEE (\ref{HEE}) in the presence of DGP gravity. Recall that cone holography proposes that the CFT on the defect $D$ is dual to the gravity in bulk coupled to a codim-2 brane $E$ and a codim-1 brane $Q$. Thus, we have $\log Z_{\text{CFT}}=- I_{\text{gravity}}$, where $Z_{\text{CFT}}$ is the CFT partition function and $I_{\text{gravity}}$ is the Euclidean bulk action (\ref{action}) including contributions from the two branes $E$ and $Q$. Since there is no DGP gravity on the codim-2 brane $E$, the brane $E$ does not modify the RT formula \cite{Jensen:2013lxa}. Thus only the DGP gravity on the codim-1 brane $Q$ makes nontrivial contributions to the entropy formula. By applying the approach of \cite{Lewkowycz:2013nqa, Dong:2013qoa}, \cite{Chen:2020uac} derives the RT formula (\ref{HEE}) in the presence of dynamical gravity on the codim-1 brane. Besides, \cite{Chen:2020uac} also makes nontrivial tests for this entropy formula. Our case of DGP cone holography is similar. Now we finish the explanation of the HEE (\ref{HEE}) for DGP cone holography.

\subsubsection{The whole space}
 Let us first discuss the HEE of the vacuum state on the whole defect $D$. To have zero HEE of this pure state \footnote{ In fact, we can relax the constraint that the HEE of the entire space is bounded from below, which gives the same bound of $\lambda$. Note that we are studying regularized finite HEE since the branes locate at a finite place instead of infinity in wedge/cone holography. Similar to Casimir energy, the regularized HEE can be negative in principle. }, we obtain a lower bound of the DGP parameter, which is stronger than the constraint (\ref{boundDGP1}) from the positivity of effective Newton's constant. 
 
Substituting the embedding functions $z=z(r)$ and $t=\text{constant}$ into the AdS metric (\ref{metric}) and entropy formula (\ref{HEE}), i.e., $S_{\text{HEE}}=4\pi A$, we get the area functional of RT surfaces
\begin{eqnarray}\label{sect2.3.1:area}
\frac{A}{2\pi}=\int_{0}^{\rho} dr\frac{\sinh(r)\cosh^{d-3}(r)}{z(r)^{d-3}} \sqrt{1+\frac{\cosh^{2}(r) z'(r)^2}{z(r)^{2}}}+\frac{2\lambda \sinh(\rho)\cosh^{d-3}(\rho)}{z^{d-3}(\rho)},
\end{eqnarray}
where $z(\rho)$ denotes the endpoint on the codim-1 brane $Q$. For simplicity, we set the horizontal volume $V=\int d^{d-3}y=1$ in this paper. From (\ref{sect2.3.1:area}), we derive the Euler-Lagrange equation
\begin{eqnarray}\label{sect2.3.1:ELEOM}
&&(d-3) z^3 \sinh (r)+(d-4) z \sinh (r) \cosh ^2(r) \left(z'\right)^2\nonumber\\
&&+\cosh ^3(r) \left(z'\right)^3 \left((d-2) \sinh ^2(r)+\cosh ^2(r)\right)\nonumber\\
&&+\frac{1}{2} z^2 \cosh (r) \left(z' (d \cosh (2 r)-d+2)+\sinh (2 r) z''\right)=0,
\end{eqnarray}
and NBC on the codim-1 brane $Q$
\begin{eqnarray}\label{sect2.3.1:NBC}
\frac{\cosh ^2(\rho ) z'(\rho )}{\sqrt{\cosh ^2(\rho ) z'(\rho )^2+z(\rho )^2}}-2 (d-3) \lambda=0.
\end{eqnarray}
Similarly, we can derive NBC on the codim-2 brane $E$
\begin{eqnarray}\label{sect2.3.1:NBCE0}
\lim_{r\to 0}\ \sinh (r) \frac{\cosh ^{d-1}(r) z(r)^{1-d} z'(r)}{\sqrt{\frac{\cosh ^2(r) z'(r)^2}{z(r)^2}+1}}=0,
\end{eqnarray}
which is satisfied automatically due to the factor $\sinh (r)$. It seems that $z'(0)$ can take any value since it always obeys the BC (\ref{sect2.3.1:NBCE0}). However, this is not the case. Solving EOM (\ref{sect2.3.1:ELEOM}) perturbatively near $r=0$, we get 
\begin{eqnarray}\label{sect2.3.1:NBCE}
z'(0)=0,
\end{eqnarray}
which means the RT surface must end orthogonally on the codim-2 brane $E$. We remark that, 
unlike wedge holography, $r=0$ is no longer a solution to cone holography.

Note that the AdS metric (\ref{metric}) is invariant under the rescale $z\to c z$. 
Due to this rescale invariance, if $z=z_0(r)$ is an extremal surface, so does $z=c z_0(r)$. Under the rescale $z\to c z$, the area functional (\ref{sect2.3.1:area}) transforms as $A \to A/c^{d-3}$.  
Recall that the RT surface is the extremal surface with minimal area. By choosing $c\to \infty$, we get the RT surface $z=c z_0(r)\to \infty $ with zero area $A=A_0/c^{d-3}\to 0$, provided $A_0$ is positive. Here $A_0$ denotes the area of the input extremal surface $z=z_0(r)< \infty$. On the other hand, if $A_0$ is negative for sufficiently negative $\lambda$, the RT surface is given by choosing $c\to 0$ so that $A=A_0/c^{d-3}\to -\infty$. To rule out this unusual case with negative infinite entropy, we must impose a lower bound on $\lambda$.  

\begin{figure}[t]
\centering
\includegraphics[width=10cm]{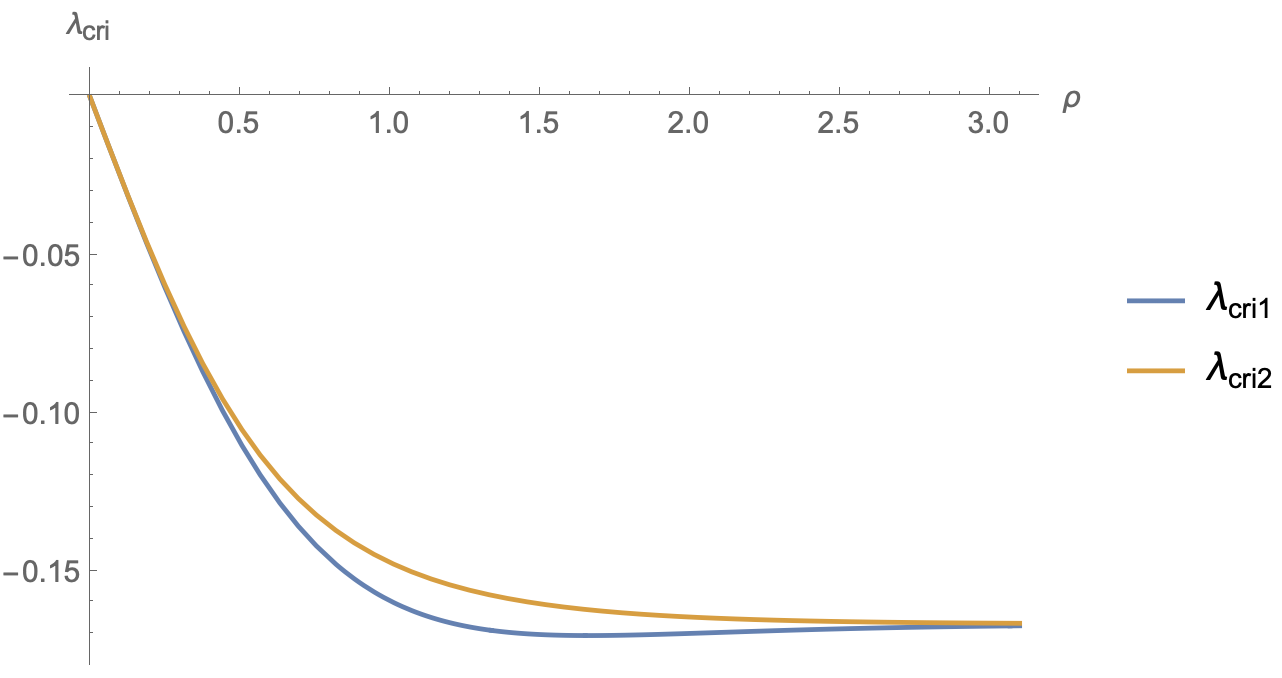}
\caption{The lower bounds of the DGP parameter for $d=5$ and $q=1$. The blue and orange curves denote the lower bounds derived from effective Newton's constant and HEE, respectively. The HEE imposes a stronger lower bound, i.e., $\lambda\ge \lambda_{\text{cri2}} \ge \lambda_{\text{cri1}}$. In the large tension limit $\rho \to \infty$, we have $ \lambda_{\text{cri1}} = \lambda_{\text{cri2}} \to -1/(2(d-2))$. }
\label{lowerbound2}
\end{figure}

The approach to derive the lower bound of $\lambda$ is as follows. We take a start point $0<z(0)<\infty$ on the codim-2 brane $E$, and impose the orthogonal condition $z'(0)=0$, then we solve EOM (\ref{sect2.3.1:ELEOM}) to determine the extremal surface $z=z_0(r)$ numerically. Next, we adjust $\lambda$ so that the area $A_0$ (\ref{sect2.3.1:area}) is non-negative. Here $\lambda$ needs not to satisfy the NBC (\ref{sect2.3.1:NBC}). As discussed above, by rescaling $z\to \lim_{c\to \infty} c z$, we get the RT surface $z=\lim_{c\to \infty} c z_0(r)\to \infty$ with vanishing area $A=\lim_{c\to \infty}A_0/c^{d-3}\to 0$. In this way, we get the lower bound of the DGP parameter
\begin{eqnarray}\label{sect2.3.1:lambdabound}
\lambda\ge \lambda_{\text{cri2}},
\end{eqnarray}
where $\lambda_{\text{cri2}}$ is derived from $A_0=0$.  Note that $A_0=0$ means that the corresponding extremal surface is the RT surface with minimal area. As a necessary condition, it should satisfy the NBCs (\ref{sect2.3.1:NBC},\ref{sect2.3.1:NBCE}) on both branes. From (\ref{sect2.3.1:NBC}), we derive
\begin{eqnarray}\label{sect2.3.1:lambdacri}
\lambda_{\text{cri2}}(\rho)=\frac{\cosh ^2(\rho ) z'(\rho )}{2 (d-3)\sqrt{\cosh ^2(\rho ) z'(\rho )^2+z(\rho )^2}},
\end{eqnarray}
where $z(\rho )$ is the endpoint of the extremal surfaces derived from arbitrary start point $z(0)$ with $z'(0)=0$.  Due to the rescale invariance of AdS, different $z(0)$ gives the same $\lambda_{\text{cri2}}$ (\ref{sect2.3.1:lambdacri}).   In other words, there are infinite zero-area RT surfaces, which obey NBCs on both branes. It is similar to the case of $\text{AdS}_3$ in AdS/BCFT and wedge holography. On the other hand, for $\lambda>\lambda_{\text{cri2}}$, the RT surface locates only at infinity, i.e., $z\to \infty$. And the NBC (\ref{sect2.3.1:NBC}) can be satisfied only at infinity for $\lambda>\lambda_{\text{cri2}}$. Please see Fig.\ref{lowerbound2} for the lower bound $\lambda_{\text{cri2}}(\rho)$, which is stronger than $\lambda_{\text{cri1}}$ (\ref{boundDGP1}) derived from the positivity of effective Newton's constant.

\subsubsection{A disk}

Let us go on to discuss HEE for a disk on the defect. The bulk metric is given 
\begin{eqnarray}\label{sect.2.3.2:metric}
ds^2=dr^2+\sinh^2(r)d\theta^2+\cosh^2(r)\frac{dz^2-dt^2+dR^2+R^2 d\Omega_{d-4}^2}{z^2}, \ 0 \le r\le \rho,
\end{eqnarray}
where $R^2\le L^2$ denotes the disk on the defect $z=0$. Substituting the embedding functions $z=z(r, R)$ and $t=\text{constant}$ into the above metric and entropy formula (\ref{HEE}), we get the area functional of the RT surface
\begin{eqnarray}\label{sect2.3.2:area}
\frac{A}{2\pi V_{S_{d-4}}}&=&\int_{\Gamma} dr dR \frac{\sinh(r) R^{d-4}\cosh^{d-3}(r)}{z^{d-3}} \sqrt{1+(\partial_Rz)^2+\frac{\cosh^{2}(r)}{z^{2}} (\partial_rz)^2} \nonumber\\
&&+\int_{\Gamma \cap Q} dR\frac{2\lambda \sinh(\rho) R^{d-4}\cosh^{d-3}(\rho)}{z^{d-3}(\rho, R)} \sqrt{1+(\partial_Rz(\rho,R))^2},
\end{eqnarray}
where $V_{S_{d-4}}$ denotes the volume of unit sphere $S_{d-4}$. From the above area functional, we derive NBC on the boundary $r=\rho$
\begin{eqnarray}\label{sect2.3.2:NBC}
&&\frac{2 \lambda \left(z^{(0,1)}(\rho ,R)^2+1\right) \left((d-4) z(\rho ,R) z^{(0,1)}(\rho ,R)+(d-3) R\right)-2 \lambda R z(\rho ,R) z^{(0,2)}(\rho ,R)}{\left(z^{(0,1)}(\rho ,R)^2+1\right)^{3/2}}\nonumber\\
&=&\frac{R \cosh ^2(\rho ) z^{(1,0)}(\rho ,R)}{z(\rho ,R) \sqrt{z^{(0,1)}(\rho ,R)^2+\frac{\cosh ^2(\rho ) z^{(1,0)}(\rho ,R)^2}{z(\rho ,R)^2}+1}}.
\end{eqnarray}
Generally, it is difficult to derive the RT surface obeying the above complicated NBC. Since the disk is symmetrical, we can make a natural guess. Inspired by \cite{Akal:2020wfl}, we find that 
\begin{eqnarray}\label{sect2.3.2:RT}
z(r,R)=\sqrt{L^2-R^2},
\end{eqnarray}
is the right RT surface satisfying both EOM and NBC (\ref{sect2.3.2:NBC}). Interestingly, the RT surface (\ref{sect2.3.2:RT}) is independent of $\lambda$. Substituting (\ref{sect2.3.2:RT}) into (\ref{sect2.3.2:area}) and noting that $S_{\text{HEE}}=4\pi A$, we derive
\begin{eqnarray}\label{sect2.3.2:area1}
S_{\text{HEE}}&=&8\pi^2\Big( \int_0^{\rho} \sinh(r)\cosh^{d-3}(r) dr+2\lambda \sinh(\rho) \cosh^{d-3}(\rho) \Big) V_{S_{d-4}}\int_0^L dR L R^{d-4} \left(L^2-R^2\right)^{1-\frac{d}{2}}\nonumber\\
&=&\frac{1}{4 G^{(d-1)}_{\text{eff}}} V_{S_{d-4}}\int_0^L dR L R^{d-4} \left(L^2-R^2\right)^{1-\frac{d}{2}},
\end{eqnarray}
which takes the same expression as the HEE of a disk in AdS$_{d-1}$/CFT$_{d-2}$. The only difference is that Newton's constant is replaced with the effective one (\ref{Newton's constant1}). It shows that the vacuum has similar entanglement properties as AdS/CFT. It is a support for cone holography with DGP gravity. 

\begin{figure}[t]
\centering
\includegraphics[width=7.5cm]{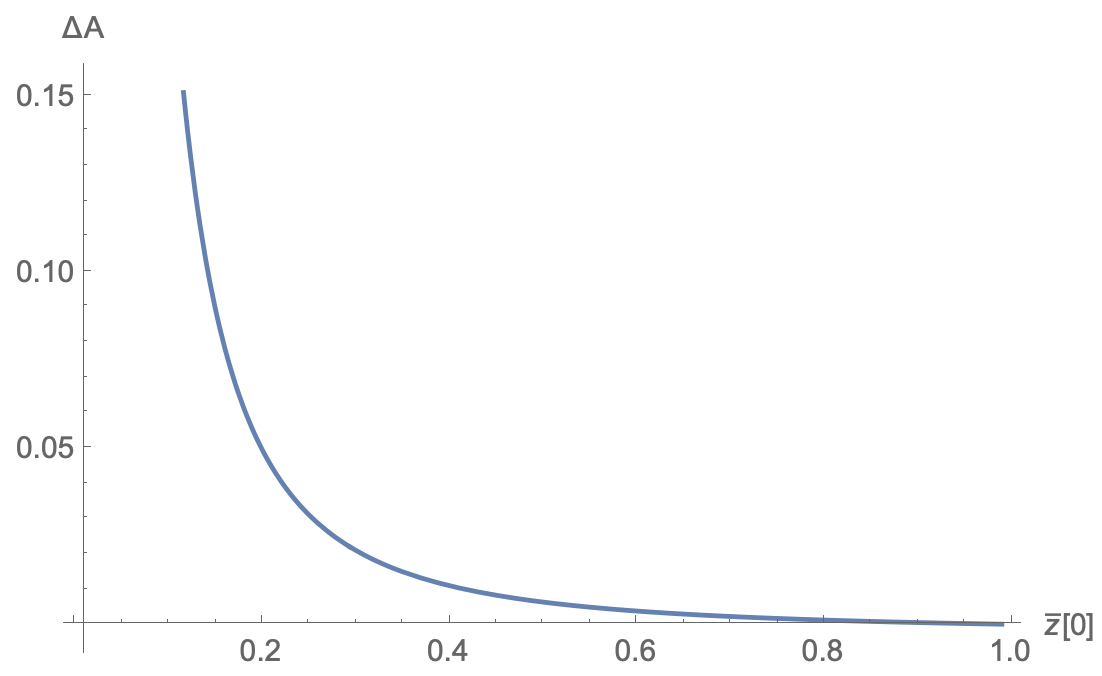}
\includegraphics[width=7.5cm]{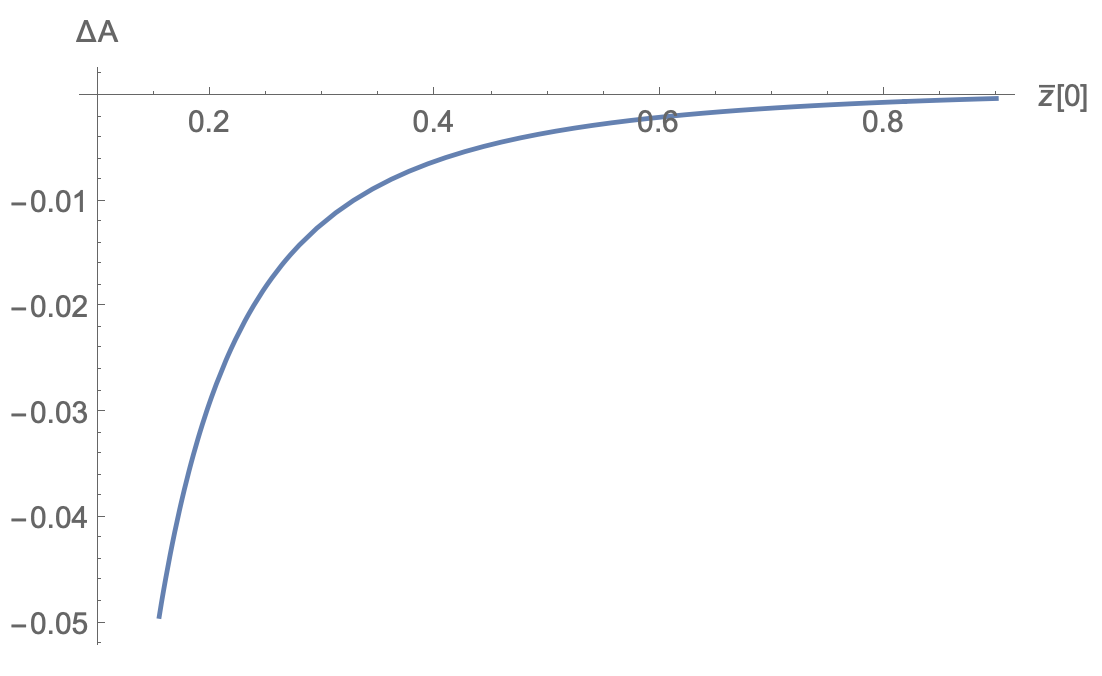}
\caption{ $\Delta A$ decreases with $\bar{z}(0)$ for $\lambda> \lambda_{\text{cri2}}$ (Left); $\Delta A$ increases with $\bar{z}(0)$ for $\lambda< \lambda_{\text{cri2}}$ (right). Here $\Delta A$ denotes the area difference between the extremal surface and the horizon of (\ref{sect.2.3.2:metricHyperbolic}), $\bar{z}(0)$ is the endpoint of the extremal surface on the codim-2 brane $E$. We choose $d=5, \rho=0.5$, which yields $\lambda_{\text{cri2}}\approx -0.104$. We choose $\lambda=-0.103> \lambda_{\text{cri2}}$ for left figure, and  $\lambda=-0.105<\lambda_{\text{cri2}}$ for right figure. It shows that the horizon area is minimal for $\lambda> \lambda_{\text{cri2}}$, while is maximum for $\lambda< \lambda_{\text{cri2}}$. }
\label{dAupdown}
\end{figure}

Recall that $\lambda$ is arbitrary in the above discussions. Now let us discuss the constraints of $\lambda$. First, we require the HEE of a disk to be positive, which yields $G^{(d-1)}_{\text{eff}}\ge 0$ and the corresponding lower bound $\lambda\ge \lambda_{\text{cri1}}$ (\ref{lowerbound1}). Second, above, we only prove (\ref{sect2.3.2:RT}) is an extremal surface obeying the NBC (\ref{sect2.3.2:NBC}). To be an RT surface, we further require that (\ref{sect2.3.2:RT}) is minimal. Remarkably, we numerically observe that this requirement yields the second lower bound $\lambda\ge \lambda_{\text{cri2}}$ (\ref{lowerbound2}). To see this, we rewrite the metric (\ref{sect.2.3.2:metric}) into the following form 
\begin{eqnarray}\label{sect.2.3.2:metricHyperbolic}
ds^2=dr^2+\sinh^2(r)d\theta^2+\cosh^2(r)\frac{\frac{d\bar{z}^2}{1-\bar{z}^2}-(1-\bar{z}^2)d\bar{t}^2+ dH_{d-3}^2}{\bar{z}^2}, \qquad 0 \le r\le \rho,
\end{eqnarray}
where $dH_{d-3}^2=dx^2+\sinh^2(x) d\Omega_{d-3}^4$ is the line element of $(d-3)$-dimensional hyperbolic space with unit curvature. Now the extremal surface (\ref{sect2.3.2:RT}) has been mapped to the horizon $\bar{z}=1$ of the hyperbolic black hole, where we have rescaled the position of the horizon. Now the problem becomes a simpler one: to find a lower bound of $\lambda$ so that the horizon $\bar{z}=1$ is the RT surface with the minimal ``area" \footnote{By ``area," we take into account the contributions from $\lambda$. }. For any given $\lambda$, we can numerically solve a class of extremal surfaces with $0<\bar{z}(0)<1$, where $\bar{z}(0)$ is the endpoint of the extremal surface on the codim-2 brane $E$. We numerically find that the horizon $\bar{z}=1$ always has the minimal area for $\lambda\ge \lambda_{\text{cri2}}$. On the other hand, the horizon area becomes maximum for $\lambda< \lambda_{\text{cri2}}$. 
Please see Fig. \ref{dAupdown}, where we take $\rho=0.5$ with $\lambda_{\text{cri2}}\approx -0.104$ as an example.

\section{Page curve for tensionless case}

\begin{figure}[t]
\centering
\includegraphics[width=8cm]{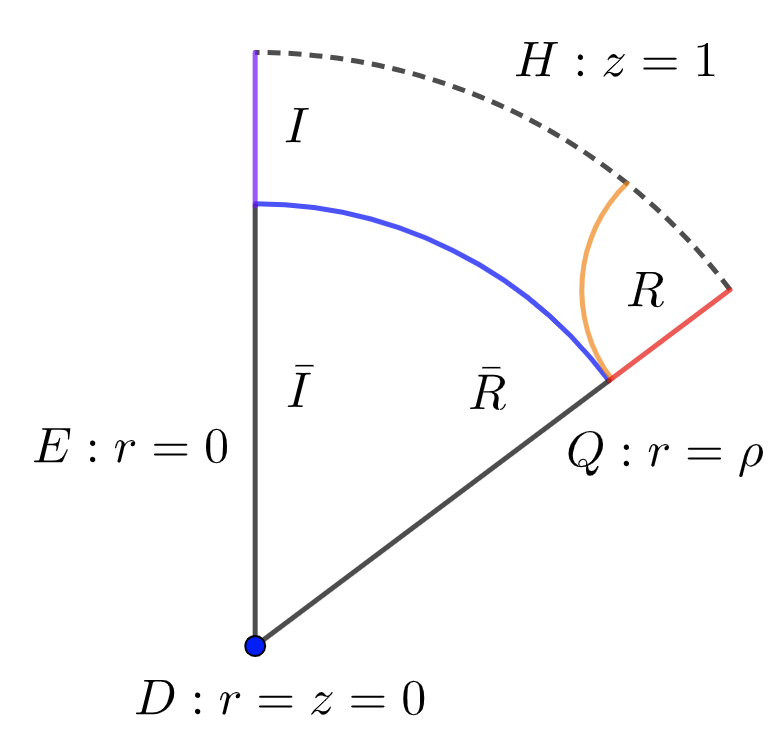}
\caption{Cone holography and its interpretations in black hole information paradox. We focus on constant angle $\theta$ and time $t$. $E$ denotes the codim-2 brane with intense gravity, and $Q$ is codim-1 brane with weak gravity. The purple and black lines denotes the island $\text{I}$ and its complement $\bar{\text{I}}$ on brane $E$, the red and black lines denotes the radiation $\text{R}$ and its complement $\bar{\text{R}}$ on brane $Q$. The dotted line, blue, and orange lines in the bulk indicate the horizon, 
the RT surface in the island phase and the HM in the no-island phase at $t=0$, respectively. }
\label{coneholographyBH}
\end{figure}

In this section, we study the information problem for eternal black holes \cite{Maldacena:2001kr} in cone holography with DGP gravity on the brane (DGP cone holography). To warm up, we focus on tensionless codim-2 branes and leave the discussion of the tensive case to the next section. See Fig. \ref{coneholographyBH} for the geometry of cone holography and its interpretations in the black hole information paradox. According to \cite{Geng:2020fxl}, since both branes are gravitating in cone holography, one should adjust both the radiation region $\text{R}$ (red line) and the island region $\text{I}$ (purple line) to minimize the entanglement entropy of Hawking radiation. Moreover, from the viewpoint of bulk, since the RT surface is minimal, it is natural to adjust its intersections $\partial \text{R}$ and $\partial \text{I}$ on the two branes to minimize its area. Following this approach, we recover non-trivial entanglement islands in cone holography with suitable DGP gravity. Furthermore, we work out the parameter space allowing Page curves, which is pretty narrow.

\begin{figure}[t]
\centering
\includegraphics[width=6cm]{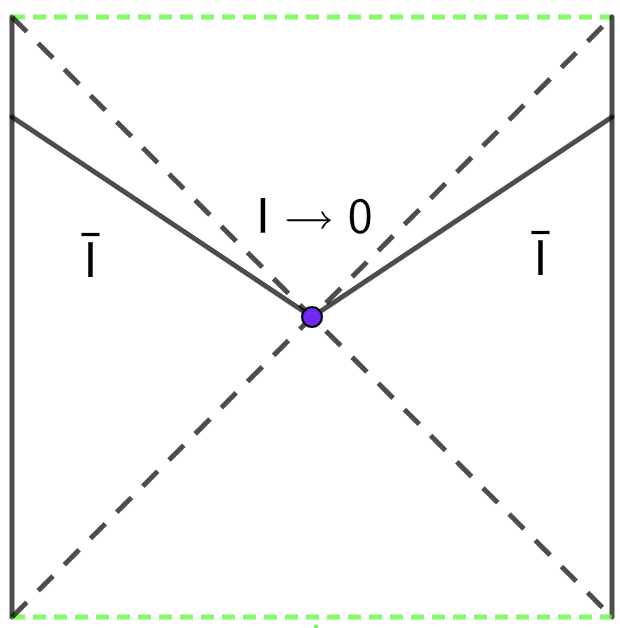}
\includegraphics[width=6cm]{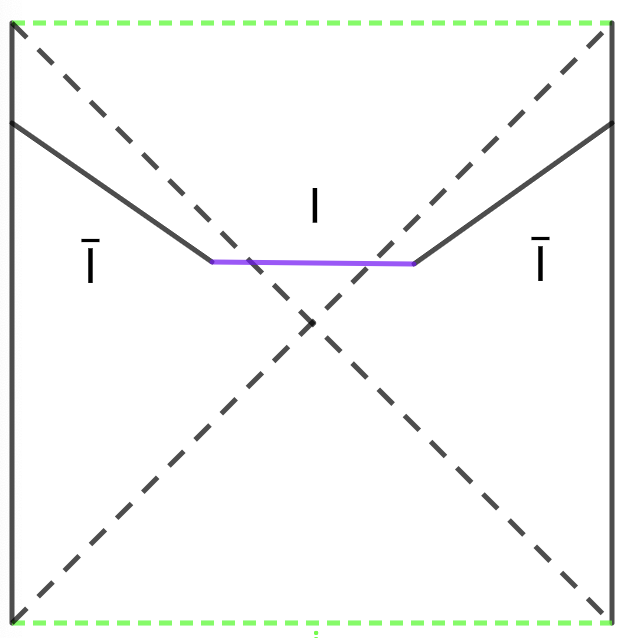}
\caption{ Left: Penrose diagram on the brane $E$ in cone holography without DGP gravity. Right: Penrose diagram on the brane $E$ in DGP cone holography. The black-dotted line, green-dotted line, and the purple line or point denote the horizon, singularity and island, respectively. It shows that the island shrinks into a point in the Penrose diagram of cone holography without DGP terms. }
\label{Penrose diagram with and without island}
\end{figure}

To start, let us explain why entanglement islands can exist in DGP cone holography. For simplicity, we focus on the black brane metric
\begin{eqnarray}\label{bhmetric}
ds^2=dr^2+\sinh^2(r)d\theta^2+\cosh^2(r)\frac{\frac{dz^2}{f(z)}-{f(z)dt^2}+\sum_{\hat{i}=1}^{d-3} dy^2_{\hat{i}}}{z^2}, \qquad 0\le r\le \rho,
\end{eqnarray}
where a black hole with $f(z)=1-z^{d-2}/z_h^{d-2}$ lives on the codim-2 brane $E$. Without loss of generality, we set $z_h=1$ below. Assuming the embedding functions $z=z(r), t=\text{constant}$ and using the entropy formula (\ref{HEE}), we obtain the area functional of RT surfaces (blue curve of Fig. \ref{coneholographyBH})
\begin{eqnarray}\label{sect3.1:area}
\frac{A_{\text{I}}}{2\pi}=\int_{0}^{\rho} dr\frac{\sinh(r)\cosh^{d-3}(r)}{z(r)^{d-3}} \sqrt{1+\frac{\cosh^{2}(r) z'(r)^2}{f(z(r))z(r)^{2}}}+\frac{2\lambda \sinh(\rho)\cosh^{d-3}(\rho)}{z(\rho)^{d-3}}.
\end{eqnarray}
where I denotes the island phase. For the case $\lambda\ge 0$, we have  
\begin{eqnarray}\label{sect3.1:noDGPAI}
\frac{A_{\text{I}}}{2\pi}\ge \int_{0}^{\rho} dr\sinh(r)\cosh^{d-3}(r)+2 \lambda \sinh(\rho)\cosh^{d-3}(\rho)=\frac{A_{\text{BH}}}{2\pi},
\end{eqnarray}
where $A_{\text{BH}}$ is the horizon area with DGP contributions, and we have used $f(z)\ge 0$ with $0\le z \le 1$. The above inequality implies the horizon $z(r)=1$ is the RT surface with minimal area for $\lambda\ge 0$. As a result, the blue curve of Fig.\ref{coneholographyBH} coincides with the horizon, and the island region $\text{I}$ (purple line) disappears \footnote{Note that the island region (purple line) envelops the black-hole horizon on the brane $E$, and only the region outside the horizon disappears.}. One can also see this from the Penrose diagram Fig.{\ref{Penrose diagram with and without island}} (left) on the brane $E$. 

Let us go on to discuss the more interesting case $\lambda <0$. For this case, the first term of (\ref{sect3.1:area}) decreases with $z(r)$, while the second term of (\ref{sect3.1:area}) increases with $z(\rho)$. These two terms compete and make it possible that there exist RT surfaces outside the horizon, i.e., $z=z(r) <1$ for sufficiently negative $\lambda$. As a result, we obtain non-trivial island regions in Fig.\ref{coneholographyBH} and Fig.{\ref{Penrose diagram with and without island}} (right). That is why we can recover entanglement islands in cone holography with negative DGP gravity. 

Recall that there are lower bounds of the DGP parameters (\ref{boundDGP1},\ref{sect2.3.1:lambdabound}). See also Fig. \ref{lowerbound2}. Therefore, we must ensure that the DGP parameter allowing islands obeys these lower bounds. It is indeed the case. Below we first take an example to recover islands and Page curves for eternal black holes and then derive the parameter space for the existence of entanglement islands and Page curves.

 \subsection{An example}

 Without loss of generality, we choose the following parameters
 \begin{eqnarray} \label{sect3.1:parameters}
 d=5, \ \rho=2,\  \lambda\approx -0.1628\approx-0.163,
\end{eqnarray}
to study the entanglement islands and Page curves.  We verify that the above DGP parameter obeys the lower bounds (\ref{boundDGP1},\ref{sect2.3.1:lambdabound})
 \begin{eqnarray} \label{sect3.1:lambdabounds}
\lambda\approx -0.163 > \lambda_{\text{cri2}}\approx  -0.165 > \lambda_{\text{cri1}}\approx  -0.170.
\end{eqnarray}

\subsubsection{Island phase}

Let us first discuss the island phase, where the RT surface ends on both branes. See the blue curve of Fig. \ref{coneholographyBH}. From the area functional (\ref{sect3.1:area}), we derive the Euler-Lagrange equation
\begin{eqnarray} \label{sect3.1:ELEOM}
z''(r) &=&-\frac{(d-6) z'(r)^2}{2 z(r)}-d \coth (2 r) z'(r)+(d-2) \text{csch}(2 r) z'(r)\notag\\
&&-\frac{z'(r)^2 \left(\coth (r) ((d-1) \cosh (2 r)+(3-d)) z'(r)+(d-2) z(r)\right)}{2 z(r)^2 f(z(r))}\notag\\
&&-2 (d-3) \tanh (r) \text{csch}(2 r) z(r) f(z(r)).
\end{eqnarray}
and NBC on the codim-1 brane $Q$
\begin{eqnarray}\label{sect3.1:NBC}
\frac{\cosh^2(\rho ) z'(\rho )}{\sqrt{\cosh^2(\rho ) z'(\rho )^2{f(z(\rho))}+z(\rho )^2 f(z(\rho))^2}}-2 (d-3) \lambda =0.
\end{eqnarray}
Similar to sect.2.3, EOM (\ref{sect3.1:ELEOM}) yields $z'(0)=0$ on the codim-2 brane $E$. By applying the shooting method, we can obtain the RT surface numerically. Let us show some details. We numerically solve EOM (\ref{sect3.1:ELEOM}) with BCs $z(0)=z_0$ and $z'(0)=0$, then we can determine $z(\rho)$ and $z'(\rho)$ on the brane $Q$. In general, $z(\rho)$ and $z'(\rho)$ does not satisfy the NBC (\ref{sect3.1:NBC}) with 
$\lambda\approx -0.163$. 
We adjust the input $z(0)=z_0$ so that the NBC (\ref{sect3.1:NBC}) is obeyed. In this way, we obtain the RT surface with two endpoints outside the horizon
\begin{eqnarray}\label{sect3.1:z0zrho}
z(0)\approx 0.963,\ z(\rho)\approx 0.900.
\end{eqnarray} 
The area of the RT surface is smaller than the horizon area (with corrections from $\lambda$)
\begin{eqnarray}\label{sect3.1:AIABH}
A_\text{I}\approx 0.694 < A_\text{BH}\approx 0.700,
\end{eqnarray} 
which verifies that there are non-trivial RT surfaces and entanglement islands outside the horizon. 

\subsubsection{No-Island phase}
Let us go on to study the RT surface in the no-island phase (HM surface, orange curve of Fig.\ref{coneholographyBH}). To avoid coordinate singularities, we choose the infalling Eddington-Finkelstein coordinate $dv=dt-\frac{dz}{f(z)}$. Substituting the embedding functions $v = v(z)$, $r = r(z)$ into the metric (\ref{bhmetric}), we get the area functional
\begin{eqnarray}\label{sect3.2:area}
\frac{A_\text{N}}{2\pi}&=&\int_{z_{\rho}}^{z_{\max}} \frac{ \sinh(r(z))\cosh^{d-3}(r(z))}{z^{d-3}} \sqrt{r'(z)^2-\frac{\cosh^2(r(z))v'(z) \left(f(z)v'(z)+2\right)}{z^2} }dz \notag\\
&& +\frac{2\lambda \sinh(\rho)\cosh^{d-3}(\rho)}{z_{\rho}^{d-3}},
\end{eqnarray}
and the time on the bath brane $Q$
\begin{eqnarray}\label{sect3.2:time}
t=t(z_{\rho}) = - \int^{z_{\max}}_{z_{\rho}} \left(v'(z)+\frac{1}{f(z)}\right)dz.
\end{eqnarray}
Here N denotes the no-island phase, $z_{\rho}$ obeying $\rho=r(z_{\rho})$ is the endpoint on the brane $Q$, $z_{\text{max}}\ge 1$ denotes the turning point of the two-side black hole. According to \cite{Carmi:2017jqz}, we have $v'(z_{\text{max}})=-\infty$ and $t(z_{\text{max}})=0$, and $z_{\text{max}}=1$ corresponds to the beginning time $t=0$.

Since the area functional (\ref{sect3.2:area}) does not depend on $v(z)$ exactly, we can derive a conserved quantity
\begin{eqnarray}\label{sect3.2:conserved quantity}
E_\text{N}=\frac{\partial L}{\partial v'(z)}=-\frac{z^{1-d} \sinh (r(z)) \left(f(z) v'(z)+1\right) \cosh ^{d-1}(r(z))}{\sqrt{r'(z)^2-\frac{\cosh ^2(r(z)) v'(z) \left(f(z)  v'(z)+2\right)}{z^2}}},
\end{eqnarray}
where $A = 2\pi  \int_{z_\rho}^{z_{\max}} L dz$. Substituting  $v'(z_{\max}) = -\infty$ and $r (z_{\max}) = r_{0}$ into the above equation, we derive
\begin{eqnarray}\label{sect3.2:conserved quantity1}
E_\text{N}&=&-\frac{z^{1-d} \sinh (r(z)) \left(f(z) v'(z)+1\right) \cosh ^{d-1}(r(z))}{\sqrt{r'(z)^2-\frac{\cosh ^2(r(z)) v'(z) \left(f(z)  v'(z)+2\right)}{z^2}}}\notag\\
&=&-\sqrt{-f(z_{\max})} \sinh \left(r_0\right) \left(z_{\max } \text{sech}\left(r_0\right)\right){}^{2-d}.
\end{eqnarray}
By applying (\ref{sect3.2:conserved quantity1}), we can delete $v'(z)$ and rewrite
 the area functional  (\ref{sect3.2:area}) and the time (\ref{sect3.2:time}) as
\begin{eqnarray}\label{sect3.2:area of r(z)} 
\frac{A_\text{N}}{2\pi}&=&\int^{z_{\text{max}}}_{z_{\rho}} dz \, \frac{\sinh(r(z))\cosh^{d-3}(r(z))}{z^{d-2}} \sqrt{\frac{\sinh ^2(r(z)) \left(z^2 f(z) r'(z)^2+\cosh ^2(r(z))\right)}{E_N^2 z^{2 d-4} \cosh ^{4-2 d}(r(z))+f(z) \sinh ^2(r(z))}}\notag\\
& & + \   \frac{2\lambda_Q\sinh(\rho)\cosh^{d-3}(\rho)}{z_\rho^{d-3}},\\
\label{sect3.2:time of r(z)} 
t&=&\int_{z_{\rho}}^{z_{\max}} \frac{z^{d-2}E_\text{N}}{f(z)}\sqrt{\frac{1+z^2f(z)\text{sech}^2(r(z))r'(z)^2}{E_\text{N}^2 z^{2d-4}+f(z)\cosh^{2d-4}(r(z))\sinh^2(r(z))}}dz.
\end{eqnarray} 
Similarly, we can simplify the EOMs derived from the area functional  (\ref{sect3.2:area}) as
 \begin{eqnarray}\label{sect3.2:ELEOM}
&&2 z^5  \cosh ^2(r )  \left(E_\text{N}^2 z^{2 d-4} \text{csch}^2(r )+ f(z) \cosh ^{2 d-4}(r )\right)  r''(z)  \notag\\
&&-2 z^3 \sinh (r ) \cosh ^{2 d-1}(r ) \left(d+\text{csch}^2(r )-1\right)+E_\text{N}^2 z^{2 d+2} \left((d-2) z^{d-2}+2 f(z)\right) r'^3 \text{csch}^2(r ) \notag\\
&&+2 E_\text{N}^2 z^{2 d} r'  \coth (r ) \left(2 \coth (r )-z r' \right) -z^4 \left(2 (d-4)  f(z)+(d-2) z^{d-2}\right) r'  \cosh ^{2 d-2}(r )  \notag\\
&&-2 z^5 f(z) r'^2 \text{csch}(r ) \left(d \sinh ^2(r )+1\right) \cosh ^{2 d-3}(r )-2 (d-3) z^6  f(z)^2  r'^3 \cosh ^{2 d-4}(r ) =0.
\end{eqnarray}
where $r=r(z)$, $r'=r'(z)$.

Solving the above equation perturbatively around the turning point, we get the BCs
\begin{eqnarray}\label{sect3.2:BCs}
r (z_{\max}) = r_{0}, \qquad \qquad \qquad  r'(z_{\max})=\frac{\coth (r_0) ((d-1) \cosh (2 r_0)-(d-3))}{ (d-2)z_{\max} \left( z_{\max}^{d-2}-2\right)}.
\end{eqnarray}
Since the RT surface ends on the bath brane $Q$, we have another BC
\begin{eqnarray}\label{sect3.2:BConQ}
r (z_{\rho}) = \rho,
\end{eqnarray}
where $\rho=2$ and $z_{\rho}\approx 0.900$ (\ref{sect3.1:z0zrho}) in our example. For any given $z_{\max}$ and $r_0$, we can  
numerically solve (\ref{sect3.2:ELEOM}) with BCs (\ref{sect3.2:BCs}), and then derive $r (z_{\rho})$. In general, $r (z_{\rho})$ does not satisfy the BC (\ref{sect3.2:BConQ}). Thus we need to adjust the input $r_0$ for given $z_{\max}$ to obey the BC (\ref{sect3.2:BConQ}). This shooting method fixes the relation between $r_0$ and $z_{\max}$ and derives $r(z)$ numerically. Substituting the numerical solution into (\ref{sect3.2:area of r(z)}) and (\ref{sect3.2:time of r(z)}), we get the time dependence of $A_{\text{N}}$ in the no-island phase.

\begin{figure}[t]
\centering
\includegraphics[width=10cm]{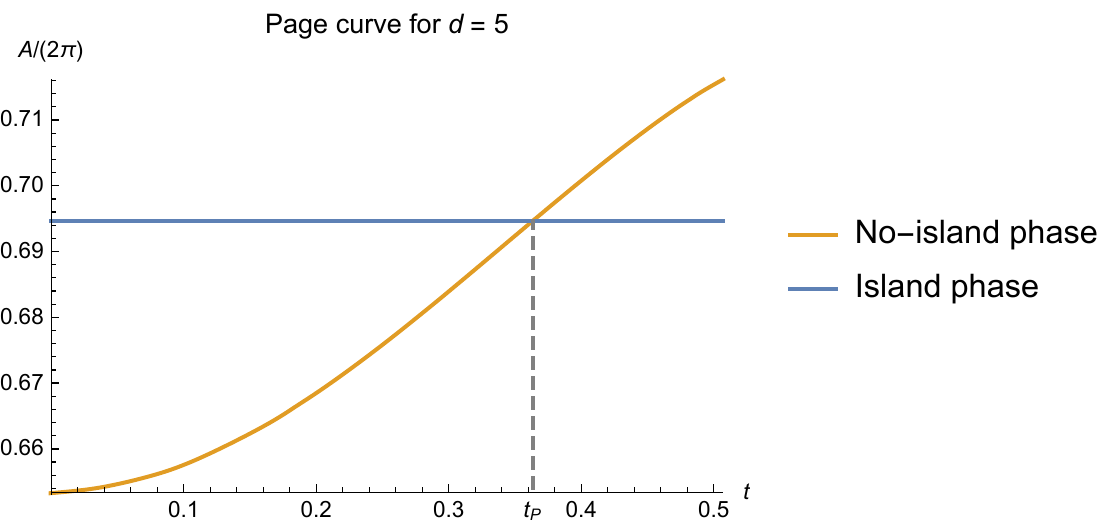}
\caption{Page curve in cone holography with DGP gravity on the codim-1 brane. Here we choose $\rho=2,\  \lambda\approx -0.163, \ d=5$, which yields the Page time $t_\text{P}\approx 0.363$. The Page curve is given by the orange line (no-island phase) for $t< t_P$ and the blue line (island phase) for $t\ge t_P$. Thus the entanglement entropy of Hawking radiation first increases with time and then becomes a constant smaller than the black hole entropy. In this way, the information paradox of the eternal black hole is resolved. }
\label{Fig:Page curve}
\end{figure}

Now we are ready to derive the Page curve. See Fig.\ref{Fig:Page curve}, where the Page curve is given by the orange line (no-island phase) for $t< t_P$ and the blue line (island phase) for $t\ge t_P$. Thus the entanglement entropy of Hawking radiation first increases with time and then becomes a constant smaller than the black hole entropy. In this way, the information paradox of the eternal black hole is resolved.

Similar to AdS/dCFT \cite{Hu:2022zgy}, the HM surface (orange line of Fig.{\ref{Fig:Page curve}}) can be defined only in a finite time. It differs from the case of AdS/BCFT and brane-world theories with only codim-1 branes. We notice that the finite-time phenomenon also appears for the HM surface of a disk in AdS/CFT. Fortunately, this unusual situation does not affect the Page curve since it happens after Page time.

\subsection{Parameter space}


In this section, we analyze the parameter space $(\rho, \lambda)$ for the existence of entanglement islands and Page curves.

\paragraph{Island Constraint 1:}  We require that the RT surfaces (blue curve of Fig.\ref{coneholographyBH}) ending on both branes locate outside the horizon, i.e., $z(r)<1$, so that there are non-vanishing island regions  (purple line of Fig.\ref{coneholographyBH}).

\begin{figure}[t]
\centering
\includegraphics[width=5cm]{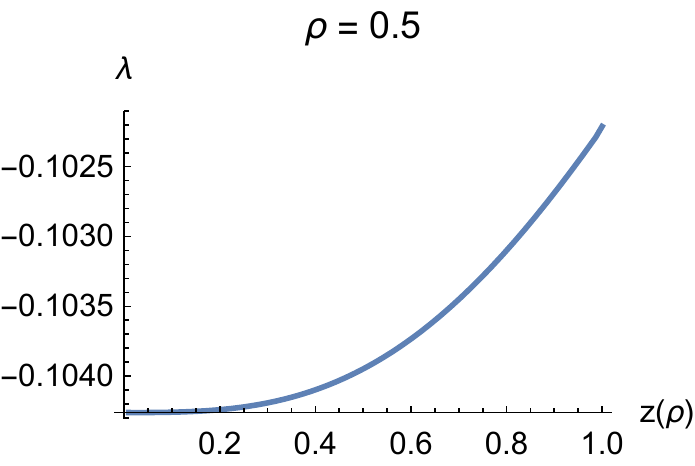}
\includegraphics[width=5cm]{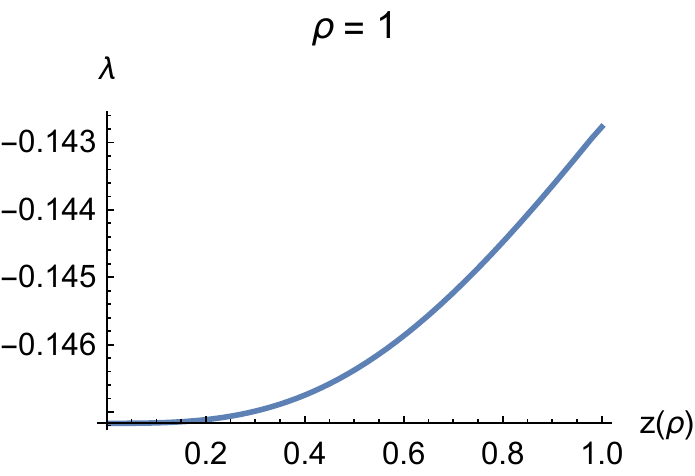}
\includegraphics[width=5cm]{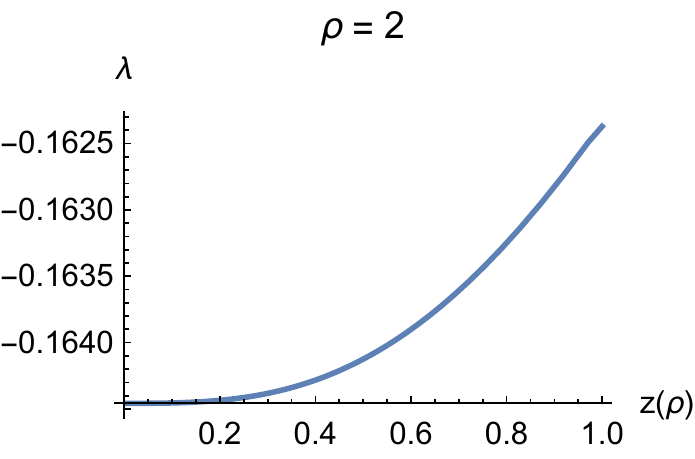}
\caption{$\lambda$ as a function of $z(\rho)$ for $\rho=0.5,1,2$ and $d=5$. Entanglement islands only exist within the range of $\lambda_{\min}=\lambda_{\text{cri2}}<\lambda<\lambda_{\max}$. }
\label{Fig:Island-Lambda}
\end{figure}

The approach to derive the parameter space obeying the island constraint is as follows. For any given $\rho$, we can obtain the extremal surface $z=z(r)$ by numerically solving (\ref{sect3.1:ELEOM}) with the BCs
\begin{eqnarray}\label{sect3.2:NBCE}
z(0) = z_0 ,\qquad z'(0)=0,
\end{eqnarray}
on the codim-2 brane $E$. The extremal surface should satisfy NBCs on both branes to become an RT surface with minimal area. From the NBC (\ref{sect3.1:NBC}) on the codim-1 brane $Q$, we derive
\begin{eqnarray}\label{sect3.2:lambda}
\lambda(\rho)= \frac{\cosh ^2(\rho ) z'(\rho )}{2(d-3)\sqrt{\cosh ^2(\rho ) z'(\rho )^2f(z(\rho))+z(\rho )^2 f(z(\rho))^2}}.
\end{eqnarray}
The above $\lambda(\rho)$ depends on the input endpoint $z_0$ on the brane $E$. Changing the endpoint from the AdS boundary $z_0=0_+$ to the horizon $z_0=1_-$, we cover all possible island surfaces outside the horizon and get the range of $\lambda$
\begin{eqnarray}\label{sect3.2:lambda allow island}
\lim_{z_0\to 0_+} \lambda(\rho)<\lambda<\lim_{z_0\to 1_-} \lambda(\rho).
\end{eqnarray}
In the limit $z\to 0_+$, the bulk geometry becomes asymptotically AdS. Thus the lower bound approaches to $\lambda_{\text{cri2}}$ (\ref{sect2.3.1:lambdacri}) in AdS 
\begin{eqnarray}\label{sect3.2:lambda allow island1}
\lim_{z_0\to 0_+} \lambda(\rho)=\lambda_{\text{cri2}}.
\end{eqnarray}
Since we have $z'(\rho)=0$ on the horizon $z=1$, one may expect that the upper bound $\lim_{z_0\to 1_-} \lambda(\rho)$ is zero. 
Remarkably, this is not the case. Although we have $z'(\rho)\to0$ and $f(z)\to 0$ near the horizon, the rate $z'(\rho)/f(z(\rho))$ is non-zero. As a result, the upper bound $\lim_{z_0\to 1_-} \lambda(\rho)$ is non-zero. Based on the above discussions, we rewrite (\ref{sect3.2:lambda allow island}) as
\begin{eqnarray}\label{sect3.2:lambda allow island3}
\lambda_{\text{cri2}}<\lambda< \lambda_{\text{max}},
\end{eqnarray}
where $ \lambda_{\text{max}}=\lim_{z_0\to 1_-} \lambda(\rho)<0.$  Take $\rho=2, d=5$ as an example, we have 
\begin{eqnarray}\label{sect3.2:lambda allow island4}
-0.1645<\lambda<-0.1623.
\end{eqnarray}
For general cases, we draw the range of $\lambda$ allowing islands as a function of $z(\rho)$ in Fig.\ref{Fig:Island-Lambda}, where one can read off the lower and upper bound of $\lambda$. Here $0<z(\rho)<1$ is the endpoint of the RT surface on the brane $Q$. From (\ref{sect3.2:lambda allow island4}) and Fig.\ref{Fig:Island-Lambda}, we see that the parameter space for the existence of entanglement islands is relatively small.

\paragraph{HM Constraint 2:}   We require that there are HM surfaces (orange curve of Fig.\ref{coneholographyBH}) ending on both the horizon and the codim-1 brane $Q$ at the beginning time $t=0$.

\begin{figure}[t]
\centering
\includegraphics[width=10cm]{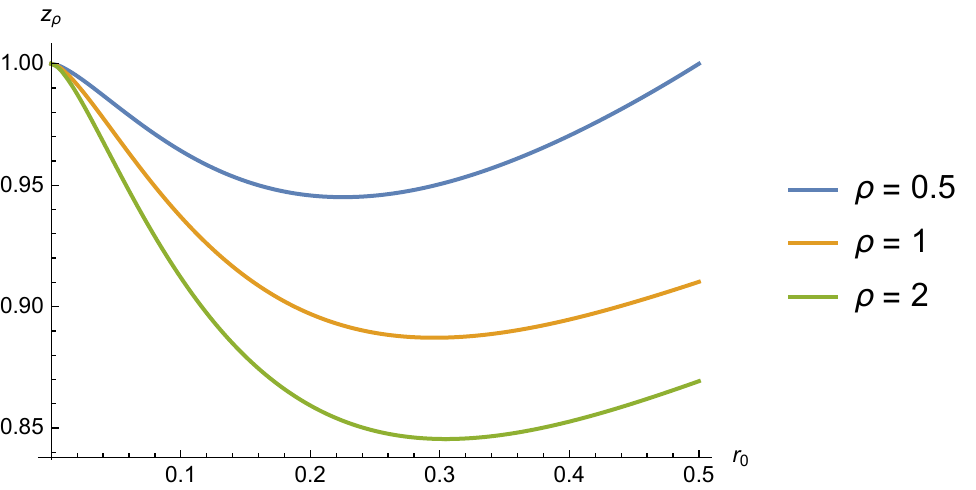}
\caption{$z_{\rho}$ as a function of $r_0$  for $d=5$ and various $\rho$, where $z_{\rho}=z(r=\rho)$ is the endpoint on the brane $Q$, and $r_0=r(z=1)$ corresponds to the endpoint on the horizon. It shows that there are lower bounds of $z(\rho)=z_{\text{cri1}}$, which yields stronger lower bounds of $\lambda$ according to Fig. {\ref{Fig:Island-Lambda}}.  }
\label{Fig:zrho-r0}
\end{figure}

\begin{figure}[t]
\centering
\includegraphics[width=14cm]{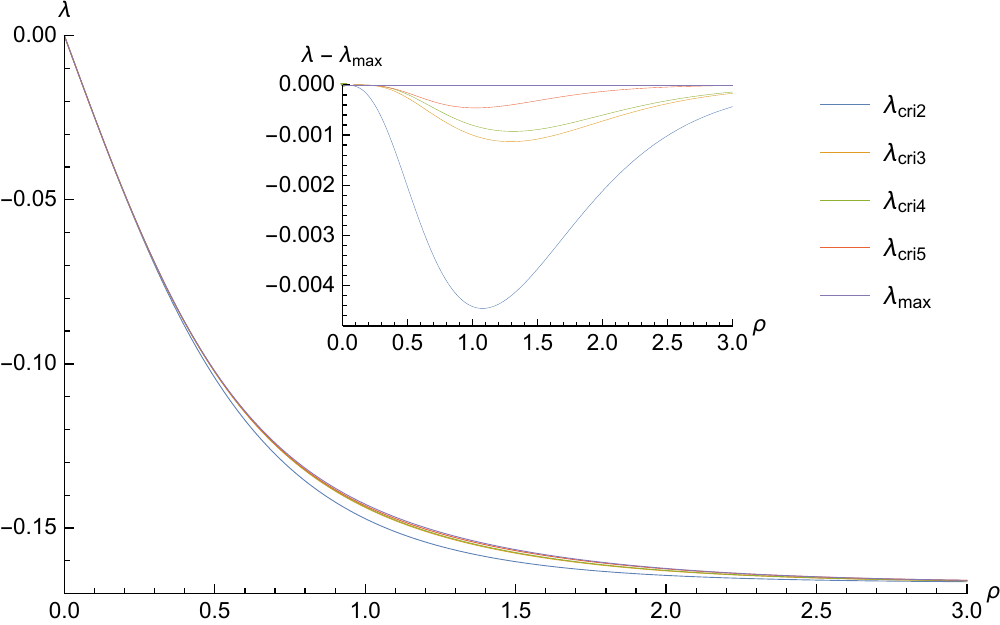}
\caption{ Various bounds of the DGP parameter $\lambda(\rho)$ for tensionless case with $d=5$. Island Constraint 1 yields $\lambda_\text{cri2} < \lambda <\lambda_\text{max}$, HM Constraint 2 gives $\lambda_\text{cri3} \leq \lambda <\lambda_\text{max}$, Page-curve Constraint 3 imposes $\lambda_\text{cri4} < \lambda < \lambda_\text{max}$ and Positive-Entropy Constraint 4 results in $\lambda_\text{cri4} < \lambda \leq \lambda_\text{cri5}$. In general we have $\lambda_\text{cri2}<\lambda_\text{cri3}<\lambda_\text{cri4} < \lambda \leq \lambda_\text{cri5}< \lambda_\text{max}$.}
\label{Fig:lambda-rho}
\end{figure}

Similar to the case in AdS/dCFT with $\rho\to \infty$ \cite{Hu:2022zgy}, HM surfaces impose a lower bound on the endpoint $z(\rho)$ in cone holography with finite $\rho$. This is quite different from the case in AdS/BCFT. Following the approach of \cite{Hu:2022zgy}, we draw $z_{\rho}=z(r=\rho)$ as a function of $r_0$ in Fig.\ref{Fig:zrho-r0}, where $r_0=r(z=1)$ denotes the endpoint of the RT surface on the horizon. Fig.\ref{Fig:zrho-r0} shows that $z(\rho)$ has a lower bound, i.e., $z(\rho)\ge z_{\text{cri1}}$. From Fig. {\ref{Fig:Island-Lambda}}, the lower bound of $z(\rho)$ produces a stronger lower bound of $\lambda$,
\begin{eqnarray}\label{sect3.3:cri3}
\lambda_\text{cri3} \leq \lambda < \lambda_{\max},
\end{eqnarray}
where $\lambda_\text{cri3}$ is given by (\ref{sect3.2:lambda}) with $z(\rho)=z_{\text{cri1}}$, $\lambda_{\max}=\lim_{z_{\rho}\to z_0\to 1_-} \lambda(\rho)$.
 See the orange line in Fig.\ref{Fig:lambda-rho} for $\lambda_\text{cri3}$.

\paragraph{Page-Curve Constraint 3:} To have the Page curve, we require that the HM surface has a smaller area than the island surfaces at the beginning time $t=0$, i.e., $A_\text{N}(t=0)<A_\text{I}$. 
Near the horizon $z_{\rho}\to 1$, the island surface (blue curve of Fig.\ref{coneholographyBH}) coincides with the horizon, and the HM surface  (orange curve of Fig.\ref{coneholographyBH}) shrinks to zero. As a result, we have $A_\text{I}-A_\text{N}(t=0) \to A_\text{hori}>0$, where $A_\text{hori}$ denotes the horizon area without DGP corrections. Thus we always have Page curves in the near-horizon limit. The reduction of $z_{\rho}$ decreases the value of $A_\text{I}(t=0)-A_\text{N}$. The critical value  $A_\text{I}-A_\text{N}(t=0)=0$ yields a lower bound $z_{\rho}= z_\text{cri2}$, which is larger than the one of HM Constraint 2, i.e., $z_\text{cri2}> z_\text{cri1}$. From Fig. {\ref{Fig:Island-Lambda}}, the stronger lower bound of $z(\rho)> z_\text{cri2} $ produces a stronger lower bound of $\lambda$ 
\begin{eqnarray}\label{sect3.3:cri4}
\lambda_\text{cri4} < \lambda < \lambda_{\max},
\end{eqnarray}
where $\lambda_\text{cri4}$ is given by (\ref{sect3.2:lambda}) with $z(\rho)=z_{\text{cri2}}$.
 See the green line in Fig.\ref{Fig:lambda-rho} for $\lambda_\text{cri4}$.

\paragraph{Positive-Entropy Constraint 4:}  Recall that we focus on regularized entanglement entropy in this paper, which can be negative in principle (as long as it is bounded from below). However, if one requires that all entanglement entropy be positive, one gets further constraint for $\lambda$. 

Assuming Page curve exists, we have $A_\text{N}(t=0)< A_\text{I}<A_\text{BH}$. Thus we only need to require $ A_\text{N}>0$ to make all entropy positive.  Recall that HM surface shrinks to zero in the near-horizon limit $z_{\rho}\to1$. Thus only the negative DGP term contribute to $A_\text{N}(t=0)$, which yields $\lim_{z_{\rho}\to 1} A_\text{N}(t=0)<0$.  To have a positive $A_\text{N}(t=0)$, we must impose a upper bound of $z_{\rho}\le z_\text{cri3}<1$, which leads to an upper bound $ \lambda_\text{cri5}$ for $ \lambda$. Combing the above discussions, the strongest bound is given by
\begin{eqnarray}\label{sect3.3:cri5}
\lambda_\text{cri4} < \lambda \le \lambda_\text{cri5},
\end{eqnarray} 
where $\lambda_\text{cri4}$ is given by (\ref{sect3.2:lambda}) with $z(\rho)=z_{\text{cri3}}$.
See the red line in Fig.\ref{Fig:lambda-rho} for $\lambda_\text{cri5}$.  Take $\rho=2, d=5$ as an example, the strongest constraint is given by
\begin{eqnarray}\label{sect3.3:boundexample}
-0.1629< \lambda \le -0.1625.
\end{eqnarray} 

To summarize, we draw various constraints of the DGP parameter $ \lambda $ in Fig. \ref{Fig:lambda-rho}, which shows the parameter space for entanglement islands and Page curves is pretty narrow. Similarly, we can also derive the parameter space for wedge holography. Please see appendix A for an example. 

\section{Page curve for tensive case}

\begin{figure}[t]
\centering
\includegraphics[width=7.5cm]{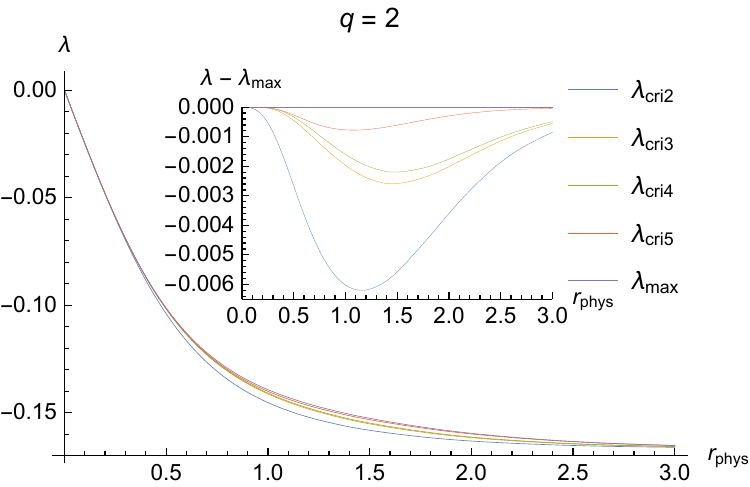}
\includegraphics[width=7.5cm]{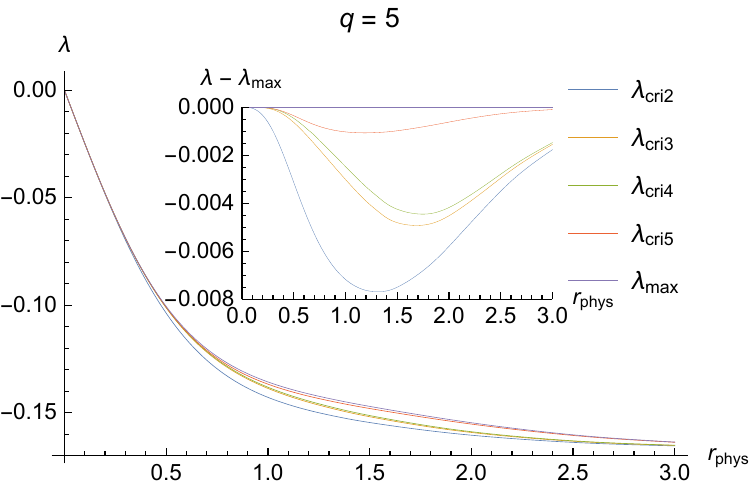}
\caption{ Various bounds of the DGP parameter $\lambda$ for $d=5$ and $q=2,5$. Here $r_{\text{phys}}$ is the physical distance between the two branes, and $r_{\text{phys}}=\rho$ in the tensionless case with $q=1$. Island Constraint 1 yields $\lambda_\text{cri2} < \lambda <\lambda_\text{max}$, HM Constraint 2 gives $\lambda_\text{cri3} \leq \lambda <\lambda_\text{max}$, Page-curve Constraint 3 imposes $\lambda_\text{cri4} < \lambda < \lambda_\text{max}$ and Positive-Entropy Constraint 4 results in $\lambda_\text{cri4} < \lambda \leq \lambda_\text{cri5}$. In general we have $\lambda_\text{cri2}<\lambda_\text{cri3}<\lambda_\text{cri4} < \lambda \leq \lambda_\text{cri5}< \lambda_\text{max}$.}
\label{Fig:lambda-rho tensive}
\end{figure}

\begin{figure}[t]
\centering
\includegraphics[width=10cm]{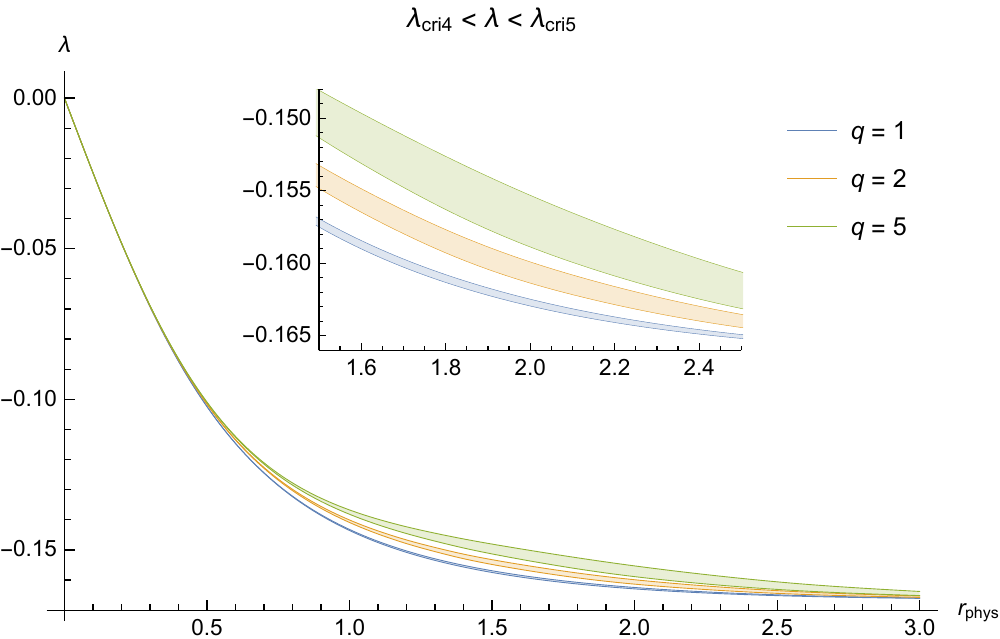}
\caption{The strongest constraint $\lambda_{\text{cri4}} < \lambda(r_{\text{phys}}) \le \lambda_{\text{cri5}}$ for $d=5$ and $q=1,2,5$. It shows that the larger the tension $q$ is, the larger the parameter space becomes.}
\label{Fig:lambda tensive}
\end{figure}

\begin{figure}[t]
\centering
\includegraphics[width=10cm]{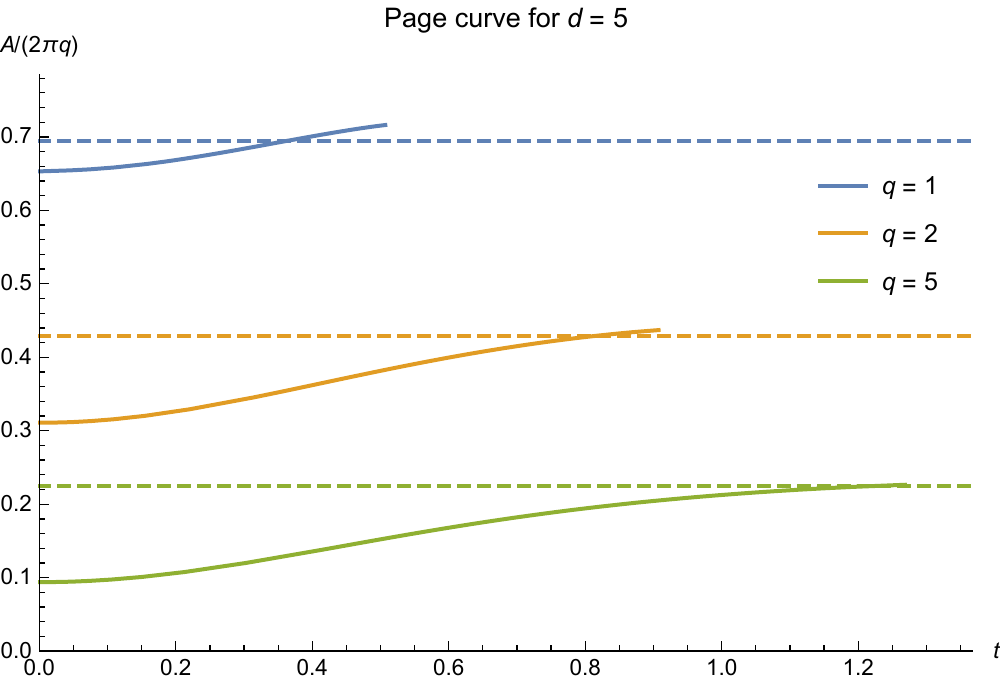}
\caption{Page curve on codim-2 brane in DGP cone holography. Here we choose $d=5$, $z_{\rho}\approx 0.900$ and $r_\text{phys}=2$. }
\label{Fig:Page curve tensive}
\end{figure}

In this section, we generalize the discussions to the case with tensive codim-2 brane $E$. Since the method is the same as sect. 3, we only show some key results below. 

The bulk metric is given by
\begin{eqnarray}\label{sect4:metric}
ds^2=\frac{d\bar{r}^2}{F(\bar{r})}+ F(\bar{r}) d\theta^2+\bar{r}^2 \frac{\frac{dz^2}{f(z)}-f(z)dt^2+\sum_{\hat{i}=1}^{d-3} dy^2_{\hat{i}}}{z^2},\qquad \bar{r}_h\le \bar{r}\le \bar{r}_0,
\end{eqnarray}
where $F(\bar{r})=\bar{r}^2-1-\frac{\bar{r}_h^{d-2}(\bar{r}_h^{2}-1)}{\bar{r}^{d-2}}$, $\bar{r}_h=\frac{1+\sqrt{d^2 q^2-2 d q^2+1}}{d q}$, $f(z)=1-z^{d-2}$. The tension of brane $E$ is given by $8\pi G_N T_E= 2\pi \left(1-\frac{1}{q} \right)$. The codim-2 brane $E$ and codim-1 brane $Q$ locate at $\bar{r}=\bar{r}_h$ and $\bar{r}=\bar{r}_0$, respectively.  The physical distance between the brane $E$ and brane $Q$ is given by
\begin{eqnarray}\label{sect4:distance}
r_{\text{phys}}=\int_{\bar{r}_h}^{\bar{r}_0} \frac{d\bar{r}}{\sqrt{F(\bar{r})}}. 
\end{eqnarray}
Below we list the EOMs and BCs used in the numeral calculations.

\paragraph{Island phase}
Substituting the embedding functions $z=z(r), t=\text{constant}$ into the metric (\ref{sect4:metric}), we get the area functional
\begin{eqnarray} \label{sect4:island Area}
\frac{A}{2\pi q} = \int_{\bar{r}_h}^{\bar{r}_\text{UV}} d\bar{r}\, \frac{\bar{r}^{d-3}}{z(\bar{r})^{d-3}} \sqrt{\frac{F(\bar{r})\bar{r}^2 z'(\bar{r})^2}{z(\bar{r})^2 (1-z(\bar{r})^{d-2})}+1 }  + \frac{2\lambda\sqrt{F(\bar{r}_0)}\bar{r}_0^{d-3}}{z(\bar{r}_0)^{d-3}}.
\end{eqnarray}
and the NBC on the codim-1 brane $Q$
\begin{eqnarray} \label{sect4:lambda}
\frac{\sqrt{F(\bar{r}_0)}\bar{r}_0^2 z'(\bar{r}_0)}{\sqrt{z(\bar{r}_0)^2 f(z(\bar{r}_0))^2+F(\bar{r}_0)\bar{r}_0^2z'(\bar{r}_0)^2f(z(\bar{r}_0))}}-2(d-3)\lambda =0 .
\end{eqnarray}
Taking variations of (\ref{sect4:island Area}), we get EOM
\begin{eqnarray} \label{sect4:island EOM}
z''(\bar{r})&=&\frac{z'(\bar{r})^2 \left(\bar{r} z'(\bar{r}) \left(2 (d-2) F(\bar{r})+\bar{r} F'(\bar{r})\right)+(d-2) z(\bar{r})\right)}{2 \left(z(\bar{r})^d-z(\bar{r})^2\right)}+\frac{(d-3) z(\bar{r})^{d-1}}{\bar{r}^2 F(\bar{r})}\notag\\
&&-\frac{(d-1) z'(\bar{r})}{\bar{r}}-\frac{(d-6) z'(\bar{r})^2}{2 z(\bar{r})}-\frac{(d-3) z(\bar{r})+\bar{r}^2 F'(\bar{r}) z'(\bar{r})}{\bar{r}^2 F(\bar{r})}.
\end{eqnarray}
Solving EOM (\ref{sect4:island EOM}) perturbatively around $\bar{r}=\bar{r}_h$, we get the BCs on the codim-2 brane $E$
\begin{eqnarray} \label{sect4:island BC}
z(\bar{r}_h)  =  z_0, \qquad z'(\bar{r}_h)=\frac{(d-3)z_0(z_0^{d-2}-1)}{\bar{r}_h(d \bar{r}_h^2-d+2)}.
\end{eqnarray}

\paragraph{No-island phase}
Substituting the embedding functions $\bar{r}=\bar{r}(z)$ and $v=v(z)$ into the metric (\ref{sect4:metric}) and defining the conserved quantity
\begin{eqnarray} \label{sect4:noisland EN}
E_\text{N2} &= &-\frac{z^{1-d}F(\bar{r}(z))\bar{r}(z)^{d-1}(f(z)v'(z)+1)}{\sqrt{\bar{r}'(z)^2-\frac{F(\bar{r}(z))\bar{r}(z)^2v'(z)(f(z)v'(z)+2)}{z^2}}}\\
&=&- z_{\max}^{2-d} \bar{r}_0^{d-2} \sqrt{-f(z_{\max})F( \bar{r}_0)} ,
\end{eqnarray}
we derive the area functional and the time on the bath brane $Q$
\begin{eqnarray} 
\frac{A}{2\pi q} &=& \int^{z_{\max}}_{z_{\bar{r}_0}} d\bar{r}\frac{\bar{r}(z)^{d-3} }{z^{d-2}}\sqrt{\frac{z^2 f(z) F(\bar{r}(z)) \bar{r}'(z)^2+\bar{r}(z)^2 F(\bar{r}(z))^2}{E_\text{N2}^2 z^{2 d-4} \bar{r}(z)^{4-2 d}+f(z) F(\bar{r}(z))}} + \frac{2\lambda\sqrt{F(\bar{r}_0)}\bar{r}_0^{d-3}}{z_{\bar{r}_0}^{d-3}},\label{sect4:no-island }\\
t &=& \int^{z_{\max}}_{z_{\bar{r}_0}} d\bar{r}\frac{E_\text{N2}z^{d-2}}{\bar{r}(z)f(z)} \sqrt{\frac{z^2 f(z) \bar{r}'(z)^2+\bar{r}(z)^2 F(\bar{r}(z))}{E_\text{N2}^2 z^{2 d-4} F(\bar{r}(z))+f(z) \bar{r}(z)^{2 d-4} F(\bar{r}(z))^2}}.
\label{sect4:time}
\end{eqnarray}
Similarly, we get the decoupled EOM for $r(z)$
\begin{eqnarray} \label{sect4:no-island EOM}
&& 2 z \bar{r}^2 F(\bar{r}) \left(E_\text{N2}^2 z^{2 d} \bar{r}^4 \bar{r}''+z^4 f(z) \bar{r}^{2 d} F(\bar{r})\right) -2 F(\bar{r}) \bar{r}' \left(E_\text{N2}^2 z^{2 d+1} \bar{r}^5 \bar{r}'-2 E_\text{N2}^2 z^{2 d} \bar{r}^6\right)  \notag\\
&& -E_\text{N2}^2 z^{2 d+1} \bar{r}^4 \bar{r}'^2 \left(z \left(z f'(z)-2 f(z)\right) \bar{r}'+\bar{r}^2 F'(\bar{r})\right)-2 (d-2) z^3 \bar{r}^{2 d+3} F(\bar{r})^3  \notag\\
&& +z^3 \bar{r}^{2 d+1} F(\bar{r})^2 \left(z \bar{r} \bar{r}' \left(z f'(z)-2 (d-4) f(z)\right)-2 (d-1) z^2 f(z) \bar{r}'^2-\bar{r}^3 F'(\bar{r})\right)  \notag\\
&& -2 F(\bar{r}) \bar{r}' \left(z^5 f(z) \bar{r}^{2 d+2} \bar{r}' F'(\bar{r})+(d-3) z^6 f(z)^2 \bar{r}^{2 d} \bar{r}'^2\right)=0.
\end{eqnarray}
and the BCs
\begin{eqnarray}\label{sect4:no-island BCs}
\bar{r} (z_{\max}) = \bar{r}_{0}, \qquad  \bar{r}'(z_{\max})= \frac{\bar{r}_0 (2(d-2)F(\bar{r}_0)+\bar{r}_0 F'(\bar{r}_0))}{(d-2)z_{\max}(z_{\max}^{d-2}-2)} .
\end{eqnarray}
Note that (\ref{sect4:no-island EOM}) is not derived from the simplified area functional (\ref{sect4:no-island }) by using the conserved quantity (\ref{sect4:noisland EN}). Instead, it is obtained from the Euler-Lagrange equation of the initial area functional, including both $r(z)$ and $v(z)$ (see (\ref{sect3.2:area}) for the tensionless case). 
Following the approach of sect.3, we derive the various bounds of the DGP parameter $\lambda$. See Table \ref{table3lambda} for $d=5, r_\text{phys}=2$. See also Fig. \ref{Fig:lambda-rho tensive} for general $r_\text{phys}$, which shows that the parameter space for the existence of entanglement islands and Page curves is quite small. The strongest constraint is given by $\lambda_\text{cri4} < \lambda \le \lambda_\text{cri5}$, which is drawn in Fig. {\ref{Fig:lambda tensive}}. It shows that the larger the tension $q$, the larger the parameter space. To end this section, we draw the Page curves for various `tension' $q$ in Fig.{\ref{Fig:Page curve tensive}}.

\begin{table}[ht]
\caption{$\lambda$ for $d=5$, $r_\text{phys}=2$ and $q=1,2,5$.}
\begin{center}
    \begin{tabular}{| c | c | c | c |  c | c | c | c| c| c|c| }
    \hline
     &  $\lambda_\text{cri1}$ &  $\lambda_\text{cri2}$ &  $\lambda_\text{cri3}$  &  $\lambda_\text{cri4}$& $\lambda$ in Fig.\ref{Fig:Page curve tensive} & $\lambda_\text{cri5}$& $\lambda_{\max}$ \\ \hline
  $q=1$    & -0.1696 & -0.1645 & -0.1631 & -0.1629& -0.1628 &-0.1625 & -0.1623 \\  \hline
  $q=2$    & -0.1733 & -0.1632 & -0.1616 & -0.1613&-0.1603&-0.1599 &-0.1596 \\ \hline
  $q=5$   & -0.1810 & -0.1605 & -0.1592 &-0.1589 &-0.1557&-0.1553 &-0.1547 \\ \hline
    \end{tabular}
\end{center}
\label{table3lambda}
\end{table}

\section{Conclusions and Discussions}

This paper investigates the information problem for eternal black holes in DGP cone holography with massless gravity on the brane. We derive the mass spectrum of gravitons and verify that there is a massless graviton on the brane. By requiring positive effective Newton's constant and zero holographic entanglement entropy for a pure state, we get two lower bounds of the DGP parameter $\lambda$. We find that entanglement islands 
exist 
in DGP cone holography obeying such bounds. Furthermore, we recover the Page curve for eternal black holes. In addition to DGP wedge holography, our work provides another example that the entanglement island is consistent with massless gravity theories. Finally, we analyze the parameter space $(\rho, \lambda)$ for the existence of entanglement islands and Page curves and find it is pretty narrow. The parameter space becomes more significant for tensive codim-2 branes. It is interesting to generalize the discussions to higher derivative gravity, such as Gauss-Bonnet gravity so that one can add non-trivial DGP gravity on the codim-2 brane. Discussing cone holography with codim-n branes and charged black holes is also enjoyable. 
In general, the quantum Hilbert space of gravity can’t be factorized. However, our results imply some approximate factorization of Hilbert space may exist at least in the semiclassical gravity approximation.  How to  factorize approximately the Hilbert space of gravity is an important question. 
We hope these issues can be addressed in the future.

\section*{Acknowledgements}

We thank J. Ren, Z. Q. Cui and Y. Guo for valuable comments and discussions. This work is supported by the National Natural Science Foundation of China (No.12275366 and No.11905297).

\appendix

\section{Parameter space of wedge holography}
\begin{figure}[t]
\centering
\includegraphics[width=14cm]{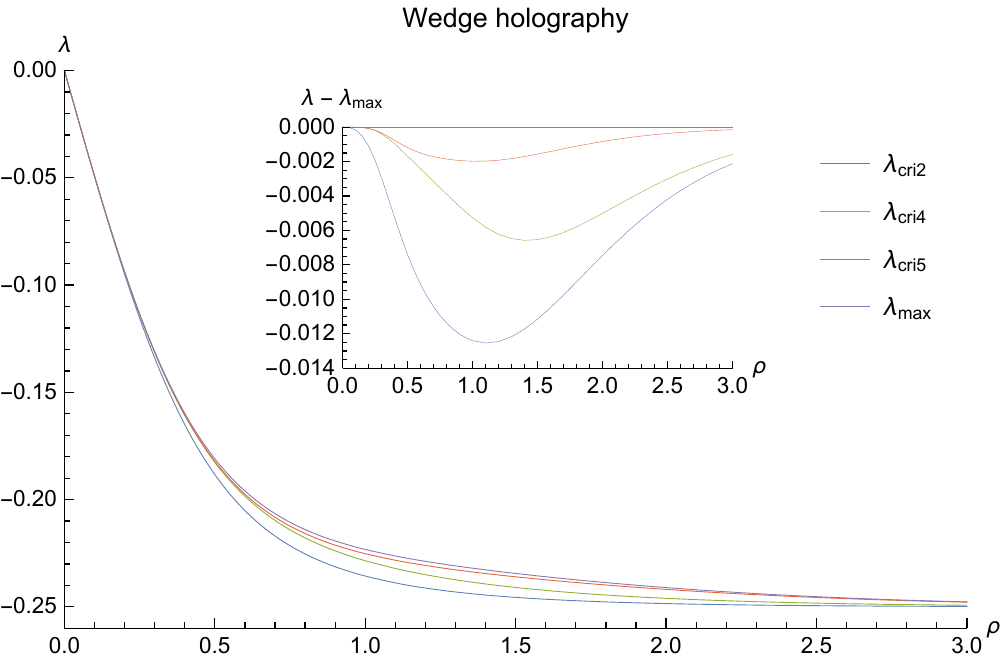}
\caption{ Various bounds of the DGP parameter $\lambda(\rho)$ for $d=4$ in wedge holography. Island Constraint 1 yields $\lambda_\text{cri2} < \lambda <\lambda_\text{max}$, Page-Curve Constraint 3 imposes $\lambda_\text{cri4} < \lambda < \lambda_\text{max}$ and Positive-Entropy Constraint 4 results in $\lambda_\text{cri4} < \lambda \leq \lambda_\text{cri5}$. In general we have $\lambda_\text{cri2}<\lambda_\text{cri4}< \lambda \leq \lambda_\text{cri5}< \lambda_\text{max}$. Note that there is not HM Constraint 2 in wedge holography. }
\label{Fig:wedge}
\end{figure}

Following the approach of sect.3.2, we can work out the parameter space for entanglement islands and Page curves for wedge holography. For simplicity, we focus on the case with $-\rho\le r\le \rho$, which corresponds to case II of \cite{Miao:2022mdx, Miao:2023unv}. Since the calculations are similar to sect.3.2, we list only the main results in this appendix. 
The parameter space is shown in Fig. \ref{Fig:wedge}, where Island Constraint 1 yields $\lambda_\text{cri2} < \lambda <\lambda_\text{max}$, Page-Curve Constraint 3 imposes $\lambda_\text{cri4} < \lambda < \lambda_\text{max}$ and Positive-Entropy Constraint 4 results in $\lambda_\text{cri4} < \lambda < \lambda_\text{cri5}$. Take $d=4$ and $\rho=0.5$ as an example, we have  
$\lambda_{\text{cri2}} \approx -0.188,  \lambda_{\text{cri4}} \approx -0.183, \lambda_{\text{cri5}} \approx -0.182 , \lambda_{\text{max}} \approx -0.181$, and the strongest constraint for the DGP parameter is
 \begin{eqnarray} \label{sectA:lambdabounds}
 \lambda_{\text{cri4}} \approx -0.183 \le \lambda \le\lambda_{\text{cri5}} \approx -0.182,
\end{eqnarray} 
which is very narrow.


\begin{thebibliography}{00}

\bibitem{Hawking:1976ra}
S.~W.~Hawking,
Phys. Rev. D \textbf{14}, 2460-2473 (1976)

\bibitem{Penington:2019npb}
G.~Penington,
JHEP \textbf{09}, 002 (2020)

\bibitem{Almheiri:2019psf}
A.~Almheiri, N.~Engelhardt, D.~Marolf and H.~Maxfield,
JHEP \textbf{12}, 063 (2019)

\bibitem{Almheiri:2019hni}
A.~Almheiri, R.~Mahajan, J.~Maldacena and Y.~Zhao,
JHEP \textbf{03}, 149 (2020)

\bibitem{Almheiri:2020cfm}
A.~Almheiri, T.~Hartman, J.~Maldacena, E.~Shaghoulian and A.~Tajdini,
Rev. Mod. Phys. \textbf{93}, no.3, 035002 (2021)
[arXiv:2006.06872 [hep-th]].


\bibitem{Karch:2000ct}
A.~Karch and L.~Randall,
JHEP \textbf{05}, 008 (2001)

\bibitem{Takayanagi:2011zk}
T.~Takayanagi,
Phys. Rev. Lett. \textbf{107}, 101602 (2011)

\bibitem{Miao:2017gyt}
R.~X.~Miao, C.~S.~Chu and W.~Z.~Guo,
Phys. Rev. D \textbf{96}, no.4, 046005 (2017)
[arXiv:1701.04275 [hep-th]].

\bibitem{Chu:2017aab}
C.~S.~Chu, R.~X.~Miao and W.~Z.~Guo,
JHEP \textbf{04}, 089 (2017)
[arXiv:1701.07202 [hep-th]].


\bibitem{Miao:2018qkc}
R.~X.~Miao,
JHEP \textbf{02}, 025 (2019)
[arXiv:1806.10777 [hep-th]].


\bibitem{Chu:2021mvq}
C.~S.~Chu and R.~X.~Miao,
JHEP \textbf{01}, 084 (2022)
[arXiv:2110.03159 [hep-th]].



\bibitem{Almheiri:2019psy}
A.~Almheiri, R.~Mahajan and J.~E.~Santos,
SciPost Phys. \textbf{9}, no.1, 001 (2020)
[arXiv:1911.09666 [hep-th]].

\bibitem{Geng:2020qvw}
H.~Geng and A.~Karch,
JHEP \textbf{09} (2020), 121
[arXiv:2006.02438 [hep-th]].


\bibitem{Chen:2020uac}
H.~Z.~Chen, R.~C.~Myers, D.~Neuenfeld, I.~A.~Reyes and J.~Sandor,
JHEP \textbf{10}, 166 (2020)
[arXiv:2006.04851 [hep-th]].


\bibitem{Ling:2020laa}
Y.~Ling, Y.~Liu and Z.~Y.~Xian,
JHEP \textbf{03}, 251 (2021)
[arXiv:2010.00037 [hep-th]].


\bibitem{Geng:2020fxl}
H.~Geng, A.~Karch, C.~Perez-Pardavila, S.~Raju, L.~Randall, M.~Riojas and S.~Shashi,
SciPost Phys. \textbf{10}, no.5, 103 (2021)
[arXiv:2012.04671 [hep-th]].


\bibitem{Geng:2021hlu}
H.~Geng, A.~Karch, C.~Perez-Pardavila, S.~Raju, L.~Randall, M.~Riojas and S.~Shashi,
JHEP \textbf{01}, 182 (2022)
[arXiv:2107.03390 [hep-th]].


\bibitem{Geng:2022fui}
H.~Geng,
``Recent Progress in Quantum Gravity: Karch-Randall Braneworld, Entanglement Islands and Graviton Mass''


\bibitem{Akal:2020wfl}
I.~Akal, Y.~Kusuki, T.~Takayanagi and Z.~Wei,
Phys. Rev. D \textbf{102}, no.12, 126007 (2020)
[arXiv:2007.06800 [hep-th]].

\bibitem{Miao:2020oey}
R.~X.~Miao,
JHEP \textbf{01}, 150 (2021)
[arXiv:2009.06263 [hep-th]].


\bibitem{Hu:2022lxl}
P.~J.~Hu and R.~X.~Miao,
JHEP \textbf{03}, 145 (2022)
[arXiv:2201.02014 [hep-th]].


\bibitem{Krishnan:2020fer}
C.~Krishnan,
JHEP \textbf{01}, 179 (2021)
[arXiv:2007.06551 [hep-th]].


\bibitem{Ghosh:2021axl}
K.~Ghosh and C.~Krishnan,
JHEP \textbf{08}, 119 (2021)
[arXiv:2103.17253 [hep-th]].

\bibitem{Yadav:2022mnv}
G.~Yadav and A.~Misra,
[arXiv:2207.04048 [hep-th]].


\bibitem{Miao:2022mdx}
R.~X.~Miao,
[arXiv:2212.07645 [hep-th]].


\bibitem{Miao:2023unv}
R.~X.~Miao,
[arXiv:2301.06285 [hep-th]].

\bibitem{Dvali:2000hr}
G.~R.~Dvali, G.~Gabadadze and M.~Porrati,
Phys. Lett. B \textbf{485}, 208-214 (2000)




\bibitem{Emparan:2023dxm}
R.~Emparan, R.~Luna, R.~Suzuki, M.~Toma\v{s}evi\'c and B.~Way,
[arXiv:2301.02587 [hep-th]].


\bibitem{Bahiru:2023zlc}
E.~Bahiru, A.~Belin, K.~Papadodimas, G.~Sarosi and N.~Vardian,
[arXiv:2301.08753 [hep-th]].














\bibitem{Rozali:2019day}
M.~Rozali, J.~Sully, M.~Van Raamsdonk, C.~Waddell and D.~Wakeham,
JHEP \textbf{05}, 004 (2020)
[arXiv:1910.12836 [hep-th]].

\bibitem{Chen:2019uhq}
H.~Z.~Chen, Z.~Fisher, J.~Hernandez, R.~C.~Myers and S.~M.~Ruan,
JHEP \textbf{03}, 152 (2020)
[arXiv:1911.03402 [hep-th]].

\bibitem{Almheiri:2019yqk}
A.~Almheiri, R.~Mahajan and J.~Maldacena,
[arXiv:1910.11077 [hep-th]].


\bibitem{Kusuki:2019hcg}
Y.~Kusuki, Y.~Suzuki, T.~Takayanagi and K.~Umemoto,
[arXiv:1912.08423 [hep-th]].

\bibitem{Balasubramanian:2020hfs}
V.~Balasubramanian, A.~Kar, O.~Parrikar, G.~Sárosi and T.~Ugajin,
[arXiv:2003.05448 [hep-th]].



\bibitem{Kawabata:2021hac}
K.~Kawabata, T.~Nishioka, Y.~Okuyama and K.~Watanabe,
JHEP \textbf{05}, 062 (2021)
[arXiv:2102.02425 [hep-th]].

\bibitem{Bhattacharya:2021jrn}
A.~Bhattacharya, A.~Bhattacharyya, P.~Nandy and A.~K.~Patra,
JHEP \textbf{05}, 135 (2021)
[arXiv:2103.15852 [hep-th]].

\bibitem{Kawabata:2021vyo}
K.~Kawabata, T.~Nishioka, Y.~Okuyama and K.~Watanabe,
[arXiv:2105.08396 [hep-th]].


\bibitem{Chen:2020hmv}
H.~Z.~Chen, R.~C.~Myers, D.~Neuenfeld, I.~A.~Reyes and J.~Sandor,
JHEP \textbf{12}, 025 (2020)
[arXiv:2010.00018 [hep-th]].





\bibitem{Bhattacharya:2021nqj}
A.~Bhattacharya, A.~Bhattacharyya, P.~Nandy and A.~K.~Patra,
[arXiv:2112.06967 [hep-th]].


\bibitem{Geng:2021mic}
H.~Geng, A.~Karch, C.~Perez-Pardavila, S.~Raju, L.~Randall, M.~Riojas and S.~Shashi,
[arXiv:2112.09132 [hep-th]].


\bibitem{Chou:2021boq}
C.~J.~Chou, H.~B.~Lao and Y.~Yang,
Phys. Rev. D \textbf{106}, no.6, 066008 (2022)
[arXiv:2111.14551 [hep-th]].

\bibitem{Ahn:2021chg}
B.~Ahn, S.~E.~Bak, H.~S.~Jeong, K.~Y.~Kim and Y.~W.~Sun,
[arXiv:2107.07444 [hep-th]].


\bibitem{Alishahiha:2020qza}
M.~Alishahiha, A.~Faraji Astaneh and A.~Naseh,
JHEP \textbf{02}, 035 (2021)
[arXiv:2005.08715 [hep-th]].


\bibitem{Gan:2022jay}
W.~C.~Gan, D.~H.~Du and F.~W.~Shu,
JHEP \textbf{07}, 020 (2022)
[arXiv:2203.06310 [hep-th]].


\bibitem{Omidi:2021opl}
F.~Omidi,
JHEP \textbf{04}, 022 (2022)
[arXiv:2112.05890 [hep-th]].

\bibitem{Hu:2022ymx}
Q.~L.~Hu, D.~Li, R.~X.~Miao and Y.~Q.~Zeng,
JHEP \textbf{09}, 037 (2022)
[arXiv:2202.03304 [hep-th]].

\bibitem{Azarnia:2021uch}
S.~Azarnia, R.~Fareghbal, A.~Naseh and H.~Zolfi,
Phys. Rev. D \textbf{104}, no.12, 126017 (2021)
[arXiv:2109.04795 [hep-th]].

\bibitem{Anous:2022wqh}
T.~Anous, M.~Meineri, P.~Pelliconi and J.~Sonner,
SciPost Phys. \textbf{13}, no.3, 075 (2022)
[arXiv:2202.11718 [hep-th]].

\bibitem{Saha:2021ohr}
A.~Saha, S.~Gangopadhyay and J.~P.~Saha,
Eur. Phys. J. C \textbf{82}, no.5, 476 (2022)
[arXiv:2109.02996 [hep-th]].



\bibitem{Geng:2022slq}
H.~Geng, A.~Karch, C.~Perez-Pardavila, S.~Raju, L.~Randall, M.~Riojas and S.~Shashi,
[arXiv:2206.04695 [hep-th]].

\bibitem{Geng:2022tfc}
H.~Geng,
[arXiv:2206.11277 [hep-th]].

\bibitem{Yu:2022xlh}
M.~H.~Yu and X.~H.~Ge,
[arXiv:2208.01943 [hep-th]].

\bibitem{Chu:2022ieq}
C.~S.~Chu and R.~X.~Miao,
[arXiv:2209.03610 [hep-th]].

\bibitem{Hu:2022zgy}
P.~J.~Hu, D.~Li and R.~X.~Miao,
JHEP \textbf{11}, 008 (2022)
[arXiv:2208.11982 [hep-th]].

\bibitem{Yadav:2023qfg}
G.~Yadav,
[arXiv:2301.06151 [hep-th]].

\bibitem{Piao:2023vgm}
Y.~S.~Piao,
[arXiv:2301.07403 [hep-th]].

\bibitem{RoyChowdhury:2022awr}
A.~Roy Chowdhury, A.~Saha and S.~Gangopadhyay,
Phys. Rev. D \textbf{106}, no.8, 086019 (2022)
[arXiv:2207.13029 [hep-th]].

\bibitem{Choudhury:2020hil}
S.~Choudhury, S.~Chowdhury, N.~Gupta, A.~Mishara, S.~P.~Selvam, S.~Panda, G.~D.~Pasquino, C.~Singha and A.~Swain,
Symmetry \textbf{13}, no.7, 1301 (2021)
[arXiv:2012.10234 [hep-th]].


\bibitem{Hung:2023mbw}
T.~N.~Hung and C.~H.~Nam,
[arXiv:2303.00348 [hep-th]].

\bibitem{Afrasiar:2023jrj}
M.~Afrasiar, J.~K.~Basak, A.~Chandra and G.~Sengupta,
[arXiv:2302.12810 [hep-th]].

\bibitem{Perez-Pardavila:2023rdz}
C.~Perez-Pardavila,
[arXiv:2302.04279 [hep-th]].

\bibitem{Basu:2022crn}
D.~Basu, Q.~Wen and S.~Zhou,
[arXiv:2211.17004 [hep-th]].

\bibitem{Kanda:2023zse}
H.~Kanda, M.~Sato, Y.~k.~Suzuki, T.~Takayanagi and Z.~Wei,
[arXiv:2302.03895 [hep-th]].



\bibitem{Miao:2021ual}
R.~X.~Miao,
Phys. Rev. D \textbf{104} (2021) no.8, 086031
[arXiv:2101.10031 [hep-th]].

\bibitem{Bostock:2003cv}
P.~Bostock, R.~Gregory, I.~Navarro and J.~Santiago,
Phys. Rev. Lett. \textbf{92}, 221601 (2004)
[arXiv:hep-th/0311074 [hep-th]].



\bibitem{Ryu:2006bv}
S.~Ryu and T.~Takayanagi,
Phys. Rev. Lett. \textbf{96}, 181602 (2006)
[arXiv:hep-th/0603001 [hep-th]].

\bibitem{Jensen:2013lxa}
K.~Jensen and A.~O'Bannon,
Phys. Rev. D \textbf{88}, no.10, 106006 (2013)
[arXiv:1309.4523 [hep-th]].

\bibitem{Lewkowycz:2013nqa}
A.~Lewkowycz and J.~Maldacena,
JHEP \textbf{08}, 090 (2013)
[arXiv:1304.4926 [hep-th]].

\bibitem{Dong:2013qoa}
X.~Dong,
JHEP \textbf{01}, 044 (2014)
[arXiv:1310.5713 [hep-th]].


\bibitem{Maldacena:2001kr}
J.~M.~Maldacena,
JHEP \textbf{04}, 021 (2003)


\bibitem{Carmi:2017jqz}
D.~Carmi, S.~Chapman, H.~Marrochio, R.~C.~Myers and S.~Sugishita,
JHEP \textbf{11}, 188 (2017)
[arXiv:1709.10184 [hep-th]].





\end{thebibliography}
\end{document}